\documentclass[journal]{IEEEtran}

\ifCLASSINFOpdf

\else

\fi

\usepackage{graphicx}
\usepackage{makecell}
\graphicspath{ {./png} }
\usepackage{amsmath}
\usepackage{algorithm}
\usepackage{algpseudocode}
\usepackage{subfigure}
\usepackage{amsfonts}
\usepackage{cite}
\usepackage{xcolor}
\usepackage{soul}
\usepackage{adjustbox}

\begin{document}

\title{Design and Prototyping of Filtering Active STAR-RIS with Adjustable Power Splitting}

\author{Rongguang~Song,
        Haifan~Yin,~\IEEEmembership{Senior Member,~IEEE,}
        Xilong~Pei, 
        Lin~Cao, 
        Taorui~Yang,
        Xue~Ren,~\IEEEmembership{Member,~IEEE}
        and~Yuanwei~Liu,~\IEEEmembership{Fellow,~IEEE}
\thanks{This work was supported by the Fundamental Research Funds for the Central Universities and the National Natural Science Foundation of China under Grant 62071191.}
\thanks{The corresponding author is Haifan Yin.}
    \thanks{R. Song, H. Yin, X. Pei, L. Cao, and T. Yang are with the School of Electronic Information and Communications, Huazhong University of Science and Technology, Wuhan 430074, China (e-mail: song\_rg@hust.edu.cn; yin@hust.edu.cn; pei@hust.edu.cn; lincao@hust.edu.cn; try@hust.edu.cn).}
    \thanks{X. Ren is with the State Key Laboratory of Radio Frequency Heterogeneous Integration, Shenzhen University, Shenzhen, China, and the College of Electronics and Information Engineering, Shenzhen University, Shenzhen, China. (e-mail: eerenxue@szu.edu.cn).}
    \thanks{Y. Liu is with the Department of Electrical and Electronic Engineering, The University of Hong Kong, Hong Kong (email: yuanwei@hku.hk).}
}

\maketitle

\begin{abstract}

Reconfigurable Intelligent Surfaces (RISs) have emerged as a transformative technology for next-generation wireless communication systems, offering unprecedented control over electromagnetic wave propagation. In particular, Simultaneously Transmitting and Reflecting RISs (STAR-RISs) have garnered significant attention due to their full-space coverage. This paper presents an active STAR-RIS, which enables independent control of both transmission and reflection phases and features out-of-band harmonic suppression. Unlike the traditional passive RIS, the proposed design integrates active amplification to overcome the inherent passive losses, significantly enhancing signal strength and system performance. Additionally, the system supports dynamic power allocation between transmission and reflection modes, providing greater flexibility to meet diverse communication demands in complex propagation environments. The versatility of the design is further validated by extending the Radar Cross Section (RCS)-based path loss model to the STAR-RIS. This design improves efficiency, flexibility, and adaptability, offering a promising solution for future wireless communication systems, particularly in scenarios requiring simultaneous control of transmission and reflection signals.

\end{abstract}


\IEEEpeerreviewmaketitle

\section{Introduction}

\IEEEPARstart{R}{econfigurable} Intelligent Surfaces (RISs) are an advanced wireless communication technology that optimizes signal transmission by intelligently controlling the propagation of electromagnetic waves. Composed of numerous programmable reflecting elements, RIS can precisely adjust the phase, amplitude, and polarization of incident signals to control their propagation paths \cite{Renzo2020smartIEEE,huang2019reconfigurable}. In 5G and upcoming 6G networks, RIS offers significant advantages: it enhances spectral efficiency by mitigating multipath fading and interference, operates with low power consumption and cost, and provides high flexibility and adaptability in dynamic environments \cite{ZTE_YUAN,wu2021intelligent}. These attributes make RISs a pivotal technology for advancing the efficiency and scalability of next-generation communication systems.

RIS systems are increasingly applied in the analog radio frequency domain, where they dynamically control the propagation path of electromagnetic waves through phase adjustment. This phase manipulation is predominantly facilitated by analog components embedded within each RIS unit, such as varactor diodes, Positive-Intrinsic-Negative (PIN) diodes, or micro-electro-mechanical systems (MEMS) switches\cite{FENG2023REVIEW}. These components enable precise control over the radiative phase of the emitted waves. Furthermore, Field Programmable Gate Arrays (FPGAs) or other control systems manage these components, allowing for seamless transitions between different operational states of RIS units. This functionality ensures effective signal control and also provides a versatile framework for optimizing electromagnetic wave propagation paths, thereby enhancing the utility of RISs in complex communication environments. \cite{Renzo2020smartIEEE,renzo2019smart}. As research in this field progresses, various RIS designs have emerged, demonstrating their important role in enabling smart wireless environments, as discussed in prior studies \cite{Renzo2020smartIEEE, renzo2019smart}. These advancements underscore the potential of RIS to significantly improve wireless communication systems under varying environmental conditions.

At present, a significant portion of RIS hardware developments predominantly concentrate on reflective-only and transmissive-only architectures, with a particular emphasis on the reflective configurations. Reflective RISs enhance signal transmission efficiency by adjusting the reflection coefficients to direct incident electromagnetic waves toward the target accurately. In contrast, transmissive RISs allow electromagnetic waves to pass through, enabling signal transmission across different spaces, thereby improving system flexibility and coverage. These two designs exhibit distinct advantages in various application scenarios: reflective RISs perform optimally when the signal source and receiver are located in the same half-space\cite{HONGWEI2024refRIS,Dai2020Ref-RIS,Pei2021Ref,yang20171600}, whereas transmissive RISs are particularly effective in complex environments where the signal source and receiver are in different half-spaces or where signals need to bypass obstacles\cite{DAI2023TRIS,huang2023TRIS,gharbieh2024TRIS}. The work \cite{Pei2021Ref} proposed a RIS composed of 1,100 varactor-based controllable elements operating at 5.8 GHz. Experimental results demonstrated that the prototype achieved a power gain of 26 dB indoors, 27 dB outdoors, and a maximum data rate of 32 Mbps over a distance of 500 meters. Another study reports on a reflective RIS based on PIN diodes, featuring 256 2-bit elements for effective beamforming at a lower cost with high gain\cite{Dai2020Ref-RIS}. The research team also developed a wireless communication prototype utilizing this RIS. Performance evaluations revealed that this RIS could achieve a high antenna gain of 21.7 dBi at 2.3 GHz and 19.1 dBi at 28.5 GHz. Compared to traditional phased arrays, the system significantly reduced power consumption. Recent research has introduced a new design for a transmitarray RIS that operates in the sub-terahertz (THz) D-band frequency range (110-170 GHz)\cite{gharbieh2024TRIS}. This design utilizes phase-change material (PCM) technology to enable precise beam control with high resolution. The 1-bit phase shift resolution is achieved by dynamically switching between the amorphous and crystalline states in the PCM-based elements. The integration of PCM technology within the unit cells significantly reduces insertion losses and power consumption, enhancing both radiation efficiency and the precision of phase control, crucial for applications in the THz spectrum.

Although reflective and transmissive RIS have demonstrated their respective advantages in certain scenarios, they exhibit limitations in complex communication environments, such as providing seamless service to both indoor and outdoor users in urban areas or enabling wide-area coverage in Unmanned Aerial Vehicle (UAV) and satellite communications. In these cases, their performance often falls short of practical deployment requirements. To address these challenges, the novel Simultaneously Transmitting and Reflecting RISs (STAR-RISs)  have been introduced\cite{Mu2021STAR-RIS,Liu2021STAR-RIS,Ahmed2023STAR-RIS,Mu2024STAR}. STAR-RISs are capable of simultaneously transmitting and reflecting signals, thereby enabling an expanded coverage area \cite{Liu2021STAR-RIS}. By independently controlling the transmission and reflection characteristics\cite{Ahmed2023STAR-RIS}, STAR-RISs significantly enhance the degrees of freedom in wireless communication systems. This not only extends the coverage area but also improves system flexibility in complex environments, particularly in scenarios involving physical obstructions. As a result, STAR-RISs represent a key advancement in the family of RIS technologies, offering a more dynamic and adaptable solution for next-generation wireless networks. In light of the numerous significant advantages of STAR-RISs, the academic community has introduced various STAR-RIS designs in recent years\cite{YANGFAN2023STAR-RIS-D,ZHANG2024STAR-RIS-D,Wu2022STAR-RIS,zeng2022STAR-RIS-D}. In \cite{YANGFAN2023STAR-RIS-D}, the design is described as a STAR-RA (Simultaneous Transmitting and Reflecting Reconfigurable Array) that employs independent beam control by integrating PIN diodes for phase control. The design enables the incident electromagnetic waves to be split into transmitted and reflected components, each with 1-bit phase control. This allows for flexible and independent beam steering in both directions, with the prototype showcasing a wide ±60° scanning range and minimal gain fluctuation, demonstrating its potential for bidirectional wireless communication applications. The research proposes a STAR-RIS, a metasurface that achieves precise 360° phase control for both transmitted and reflected waves using a single PIN diode per unit cell\cite{ZHANG2024STAR-RIS-D}. This design allows independent control of the transmitted and reflected beams at the same frequency and polarization. A 9 $\times$ 9 metasurface prototype demonstrated excellent performance, with aperture efficiencies of 32.2\% and 33.7\% for reflection and transmission, respectively. The structure offers high flexibility and cost-effectiveness, making it particularly suitable for bidirectional communication scenarios, such as aviation and relay communications. Furthermore, the emergence of STAR-RISs not only offers a significant technological solution to address the challenges posed by complex communication environments but also lays a robust theoretical and technical foundation for the deployment of future wireless communication systems across diverse applications. 

Passive RISs have the potential to enhance the performance of traditional wireless networks, yet they face significant challenges due to the ``double fading" effect. This phenomenon involves the path loss from the transmitter (TX) to the RIS and then to the receiver (RX), which is the product of the individual path losses from TX to RIS and RIS to RX. Therefore, this compounded path loss is often substantially greater than the path loss of the direct link, limiting the potential capacity gains of passive RISs in various wireless communication scenarios\cite{ZHI2022ACTIVE,YANGFAN2023ACTIVE,LONG2021ACTIVE,AHMED2024ACTIVE}. Moreover, the insertion loss introduced by RISs\cite{yang2016study} and the losses caused by amplitude-phase coupling are also critical factors that cannot be overlooked\cite{caolin2024tap}. These losses further degrade the overall system performance, especially in complex wireless propagation environments, where insertion loss can significantly reduce the gain of RISs, limiting its potential benefits. Additionally, amplitude-phase coupling effects may impair the precision of phase control, thereby degrading signal quality and adversely affecting beamforming capabilities. The current designs of active RISs structures have demonstrated significant potential in further enhancing the system gain of RISs, as widely validated in various studies\cite{wu2022ACTIVEdesign,wang2023ACTIVEdesign,rao2023ACTIVEdesign}. By integrating RIS units with power amplifier chips, active RIS achieves significant signal enhancement compared to traditional RIS designs. This approach boosts system gain, typically reaching between 7.7 and 15 dB or even higher, while also offering reconfigurable phase control and beamforming capabilities for real-time adjustment of incident electromagnetic waves. This greatly improves adaptability to complex wireless environments. Unlike traditional passive RISs, active RISs effectively mitigate signal attenuation over long-distance transmissions, while integrated power control reduces power consumption and system costs, making it highly promising for future 6G networks.

To the best of our knowledge, no active STAR-RIS architecture has been proposed in the current literature. To address this gap, this paper proposes a novel active STAR-RIS design specifically tailored for the 2.6 GHz band, which is one of the commercial 5G frequency bands used by China Mobile, incorporating power amplifiers and Single Pole Double Throw (SPDT) switches to enable both active gain and precise phase control of the RIS elements. This architecture not only significantly enhances system performance by amplifying signal gain but also provides greater flexibility in beamforming, positioning it as a promising solution for optimizing the performance of next-generation wireless communication networks. Additionally, the design offers adjustable power distribution between transmission and reflection, greatly enhancing flexibility and adaptability. Building on previous research from the team \cite{Wangzipeng2021RCS,2-bitactiveRIS}, this paper extends a path loss model based on Radar Cross Section (RCS) to active STAR-RIS-aided systems. This model provides a foundation for the performance evaluation of STAR-RIS-aided systems. The model is further validated by experiments.

The rest of this paper is organized as follows. Sec. II introduces the modeling of the STAR-RIS-aided wireless communication system. In Sec. III, an active STAR-RIS element structure is proposed. Sec. IV shows a prototype of the active STAR-RIS, and measurements of the structural design and system model are conducted.

\section{System Model}
\label{subsec:RCS}

\begin{figure}[!htbp]
    \centering
    \includegraphics[width=0.4\textwidth]{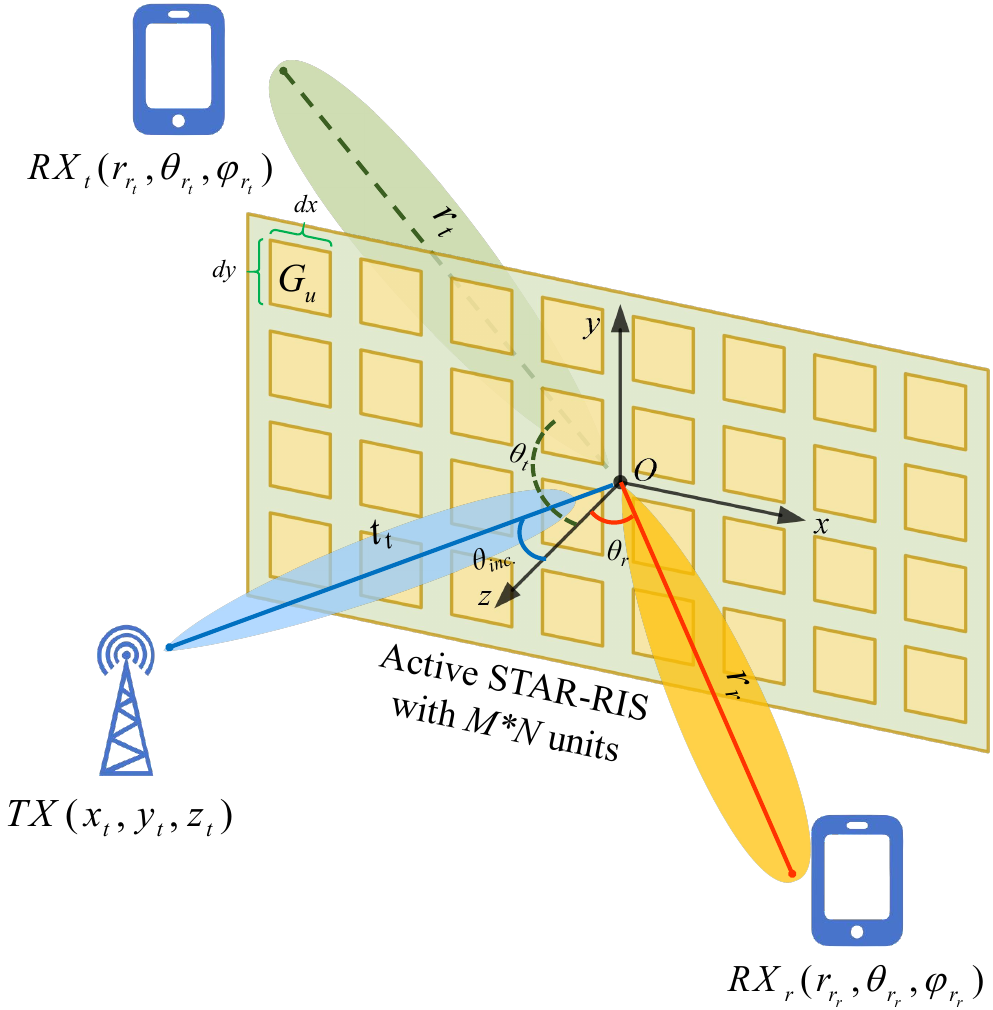}
    \caption{Geometric illustration of the active STAR-RIS-assisted wireless communication system.}
    \label{fig:system caption}
\end{figure}

We consider a wireless communication system assisted by an active STAR-RIS, which consists of a TX, a transmission receiver ($RX_t$), and a reflection receiver ($RX_r$), as shown in Fig.~\ref{fig:system caption}. The STAR-RIS is composed of $N = N_x \times N_y$ controllable elements arranged in a Uniform Planar Array (UPA) on the $xOy$ plane, with its geometric center at the origin. Each STAR-RIS element can dynamically control the power ratio between the transmitted and reflected signals. The geometric configuration of the system is as follows.

The system under consideration includes a transmitter (TX), a transmission receiver ($RX_t$), and a reflection receiver ($RX_r$). The position of the transmitter is denoted by $\boldsymbol{p}_t = (x_t, y_t, z_t)$, while the position of the transmission receiver is $\boldsymbol{p}_{r_t} = (x_{r_t}, y_{r_t}, z_{r_t})$. The reflection receiver is positioned at $\boldsymbol{p}_{r_r} = (x_{r_r}, y_{r_r}, z_{r_r})$. The STAR-RIS is composed of multiple controllable elements, each located at $\boldsymbol{p}_n = (x_n, y_n, z_n)$.

To better describe the three-dimensional positioning, we use spherical coordinates to represent the positions of the transmitter and receivers relative to the STAR-RIS. Specifically, we denote the spherical coordinates of the transmitter as $(r_t, \theta_t, \varphi_t)$, where $r_t$ is the distance from the STAR-RIS, $\theta_t$ is the zenith angle, and $\varphi_t$ is the azimuth angle. Similarly, the spherical coordinates of the transmission receiver are given by $(r_{r_t}, \theta_{r_t}, \varphi_{r_t})$, and the coordinates of the reflection receiver are represented by $(r_{r_r}, \theta_{r_r}, \varphi_{r_r})$.

The coordinates of the $n$-th STAR-RIS element can be expressed as:
\begin{equation} 
\begin{split}
\boldsymbol{p}_n = \left( \delta_{n}^{y} d_x, \delta_{n}^{x} d_y, 0 \right),
\end{split} 
\end{equation}
where $\delta_{n}^{y} = n_y - \frac{N_y + 1}{2}$ and $\delta_{n}^{x} = \frac{N_x + 1}{2} - n_x$. $d_x$ and $d_y$ represent the width and height of each STAR-RIS element, respectively.


The role of the active STAR-RIS is to shift the phase, amplitude, and power distribution between the transmitted and reflected signals, where $\eta_t$ represents the transmission power distribution coefficient, $0 < \eta_t < \eta_n$, $\eta_r$ represents the reflection power distribution coefficient, similarly satisfying $0 < \eta_r < \eta_n$. In this study, we first model the channel from the transmitter to the STAR-RIS elements. Subsequently, we model the channels from the STAR-RIS elements to the two receivers.

\subsection{Channel from Transmitter to STAR-RIS}

The channel between the transmitter (TX) and each STAR-RIS element is described by a complex channel coefficient vector $\boldsymbol{f} \in \mathbb{C}^{1 \times N}$, where the channel coefficient for the $n$-th element is $f_n$. In the free space, the channel model typically includes path loss, antenna gain, and distance-related terms. The channel coefficient for the $n$-th element is given by
\begin{equation} 
\begin{split}
f_n = \alpha_n e^{-j\xi_n},
\end{split} 
\end{equation}
where $\alpha_n$ represents the channel attenuation and $\xi_n$ represents the phase of the channel.

According to the free-space path loss model, the channel coefficient can be expressed as
\begin{equation}
\begin{split}
f_n = \sqrt{\frac{G_t\left(\boldsymbol{\hat{r}}_n^{t} \right) A_t\left(\boldsymbol{\hat{r}}_n^{t} \right)}{4 \pi}} \frac{e^{-j \frac{2 \pi}{\lambda} r_n^t}}{r_n^t},
\end{split} 
\end{equation}
where $\lambda$ is the wavelength, $G_t\left(\boldsymbol{\hat{r}}_n^{t}\right)$ is the antenna gain of the transmitter in the direction of the $n$-th element, $A_t\left(\boldsymbol{\hat{r}}_n^{t}\right)$ is the effective area of the $n$-th element in the direction of the transmitter, $r_n^t$ is the distance between the transmitter and the $n$-th element, and $e^{-j \frac{2 \pi}{\lambda} r_n^t}$ represents the phase change of the signal propagation.

\subsection{Channel from STAR-RIS to Receivers}

Similarly, the transmission receiver $RX_t$ and reflection receiver $RX_r$ are related to STAR-RIS elements through channel coefficient vectors $\boldsymbol{g}_t \in \mathbb{C}^{N \times 1}$ and $\boldsymbol{g}_r \in \mathbb{C}^{N \times 1}$, respectively. For the transmission path, the channel coefficient between the $n$-th element and $RX_t$ is given by
\begin{equation} 
\begin{split}
g_{t_n} = \beta_{t_n} e^{-j \zeta_{t_n}},
\end{split} 
\end{equation}
where $\beta_{t_n}$ represents the amplitude attenuation, and $\zeta_{t_n}$ represents the phase.

In free space, the channel model is expressed as
\begin{equation} 
\begin{split}
g_{t_n} = \sqrt{\frac{G_{r_t}\left(\boldsymbol{\hat{r}}_n^{r_t} \right) A_{r_t}\left(\boldsymbol{\hat{r}}_n^{r_t} \right)}{4\pi}}\frac{e^{-j\frac{2\pi}{\lambda}r_n^{r_t}}}{r_n^{r_t}},
\end{split}
\end{equation}
where $G_{r_t}\left(\boldsymbol{\hat{r}}_n^{r_t}\right)$ is the antenna gain of the transmission receiver, $A_{r_t}\left(\boldsymbol{\hat{r}}_n^{r_t}\right)$ is the effective area of the $n$-th element in the direction of the transmission receiver, and $r_n^{r_t}$ is the distance between the $n$-th element and $RX_t$.

Similarly, the channel coefficient between the $n$-th element and the reflection receiver $RX_r$ is given by
\begin{equation} 
\begin{split}
g_{r_n} = \beta_{r_n} e^{-j \zeta_{r_n}},
\end{split} 
\end{equation}
and in free space, it is expressed as:
\begin{equation} 
\begin{split}
g_{r_n} = \sqrt{\frac{G_{r_r}\left(\boldsymbol{\hat{r}}_n^{r_r} \right) A_{r_r}\left(\boldsymbol{\hat{r}}_n^{r_r} \right)}{4\pi}}\frac{e^{-j\frac{2\pi}{\lambda}r_n^{r_r}}}{r_n^{r_r}},
\end{split} 
\end{equation}
where $G_{r_r}\left(\boldsymbol{\hat{r}}_n^{r_r}\right)$ is the antenna gain of the reflection receiver, $A_{r_r}\left(\boldsymbol{\hat{r}}_n^{r_r}\right)$ is the effective area of the $n$-th element in the direction of the reflection receiver, and $r_n^{r_r}$ is the distance between the $n$-th element and $RX_r$.

\subsection{Received Signal Model and Power Control}

Each active STAR-RIS element has the capability to adjust the power distribution and gain dynamically; in the power distribution model of a STAR-RIS element, $\eta_t$ represents the transmission power distribution coefficient, $\eta_r$ represents the reflection power distribution coefficient, and the relationship is given by

\begin{equation}
\eta_t + \eta_r = \eta_n  \label{eq:Divider},
\end{equation}

\( \eta_n \) represents the overall radiation efficiency of the STAR-RIS unit under different power distribution ratios. It quantifies the efficiency with which the unit radiates power. This occurs when the unit is both retransmitting and reflecting an incident signal.
The received signal at $RX_t$ is
\begin{equation} 
\begin{split}
y_t = \sum_{n=1}^{N} f_n \Gamma_{t_n} g_{t_n} \sqrt{P_t} x + z_t,
\end{split} 
\end{equation}
where $f_n$ is the channel coefficient between TX and the $n$-th element, $g_{t_n}$ is the channel coefficient between the $n$-th element and $RX_t$, $\Gamma_{t_n}$ is the source coefficient, $P_t$ is the transmit power, $x$ is the transmitted signal, and $z_t \sim \mathcal{N}_{\mathbb{C}}(0, \sigma_t^2)$ is the additive white Gaussian noise (AWGN).

Considering the active gain $G_n$ of the $n$-th active STAR-RIS element, which is provided by the power amplifier, the received signal at $RX_t$ becomes
\begin{equation} 
\begin{split}
y_t = \frac{\eta_t \sqrt{P_t}}{4\pi} \sum_{n=1}^{N} \frac{\sqrt{G_t G_{r_t}  G_n A_t A_{r_t}}}{r_n^t r_n^{r_t}} \mu_{t_n} e^{j \left( \phi_{t_n} - \Phi_{t_n} \right)} x + z_t,
\end{split}
\end{equation}
where $\Phi_{t_n} = \frac{2\pi}{\lambda}(r_n^t + r_n^{r_t})$ represents the phase change due to propagation delay, The amplitude attenuation $\mu_n$ of the $n$-th element is closely related to the radiation pattern of the element.

Similarly, the received signal at $RX_r$ is
\begin{equation} 
\begin{split}
y_r = \sum_{n=1}^{N} f_n \Gamma_{r_n} g_{r_n} \sqrt{P_t} x + z_r,
\end{split}
\end{equation}
which can be further written as
\begin{equation} 
\begin{split}
y_r = \frac{\eta_r \sqrt{P_t}}{4\pi} \sum_{n=1}^{N} \frac{\sqrt{G_t G_{r_r}  G_n A_t A_{r_r}}}{r_n^t r_n^{r_r}} \mu_{r_n} e^{j \left( \phi_{r_n} - \Phi_{r_n} \right)} x + z_r,
\end{split} 
\end{equation}
where $\Phi_{r_n} = \frac{2\pi}{\lambda}(r_n^t + r_n^{r_r})$.

\subsection{RCS-based Path Loss Model}

To quantify the effect of STAR-RIS elements on signal transmission and reflection, we introduce the concept of RCS, which describes the scattering capability of STAR-RIS elements in different directions. The RCS in the transmission direction for the $n$-th element is:
\begin{equation} 
\begin{split}
\sigma_{t_n} \left( \boldsymbol{\hat{r}}_n^t, \boldsymbol{\hat{r}}_n^{r_t}, u_{t_n}\right) = \eta_t \mu_{t_n} \sqrt{\left(u_{t_n} \right) A_t\left(\boldsymbol{\hat{r}}_n^t\right) A_{r_t}\left(\boldsymbol{\hat{r}}_n^{r_t}\right)G_n} ,
\end{split}
\end{equation}
and the RCS in the reflection direction is:
\begin{equation}
\begin{split}
\sigma_{r_n} \left( \boldsymbol{\hat{r}}_n^t, \boldsymbol{\hat{r}}_n^{r_r}, u_{r_n}\right) = \eta_r \mu_{r_n} \sqrt{\left(u_{r_n} \right) A_t\left(\boldsymbol{\hat{r}}_n^t\right) A_{r_r}\left(\boldsymbol{\hat{r}}_n^{r_r}\right)G_n}.
\end{split}
\end{equation}

Thus, the received signal power at $RX_t$ and $RX_r$ can be expressed as:
\begin{equation} 
\begin{split}
P_{r_t} = \frac{P_t}{16 \pi^2} \left| \sum_{n=1}^{N} \frac{\sqrt{G_t G_{r_t} G_n}}{r_n^t r_n^{r_t}} \sigma_{t_n} e^{j \left( \phi_{t_n} - \frac{2\pi}{\lambda} \left( r_n^t + r_n^{r_t} \right) \right)} \right|^2,
\end{split}
\end{equation}
\begin{equation} 
\begin{split}
P_{r_r} = \frac{P_t}{16 \pi^2} \left| \sum_{n=1}^{N} \frac{\sqrt{G_t G_{r_r} G_n}}{r_n^t r_n^{r_r}} \sigma_{r_n} e^{j \left( \phi_{r_n} - \frac{2\pi}{\lambda} \left( r_n^t + r_n^{r_r} \right) \right)} \right|^2.
\end{split} 
\end{equation}

\subsection{Path Loss Optimization with Dynamic Power Control}

To maximize the received power, the phase shifts $\phi_{t_n}$ and $\phi_{r_n}$ applied to each STAR-RIS element, as well as the power allocation coefficients $\eta_t$ and $\eta_r$, can be adjusted so that the signals in the transmission and reflection paths coherently combine. Under ideal continuous phase shifting, the minimum path loss is given by
\begin{equation}
\begin{split}
PL_{r_t, \min} = \frac{16 \pi^2}{\left| \sum_{n=1}^{N} \frac{\sqrt{G_t G_{r_t} G_n}}{r_n^t r_n^{r_t}} \sigma_{t_n} \right|^2},
\end{split} 
\end{equation}
\begin{equation} 
\begin{split}
PL_{r_r, \min} = \frac{16 \pi^2}{\left| \sum_{n=1}^{N} \frac{\sqrt{G_t G_{r_r} G_n}}{r_n^t r_n^{r_r}} \sigma_{r_n} \right|^2}.
\end{split} 
\end{equation}

\subsection{Practical Challenges and Discrete Phase Shifting}

In practical systems, due to hardware limitations, STAR-RIS typically employs discrete phase shifting. For an $m$-bit quantized phase shifter, the available phase shift values are
\begin{equation} 
\begin{split}
\phi_n = \left\{0, \frac{\pi}{2^{m-1}}, \cdots, \frac{2^m-1}{2^{m-1}}\pi \right\}.
\end{split}
\end{equation}
Additionally, due to imperfect channel state information (CSI), the actual path loss is typically higher than the theoretical minimum. Beamforming algorithms based on discrete phase shifting are, therefore, essential for optimizing system performance.

\section{Design of STAR-RIS}

This section first provides a detailed introduction to the circuit design and implementation in the STAR-RIS architecture, with a focus on the key aspects of signal amplification, phase shifting, and power distribution circuits. To verify the performance of the circuit, we will develop an evaluation board integrated with the core circuit modules. Through experimental validation, we systematically measure and analyze the performance of the circuit, ensuring its feasibility and robustness in practical applications. Subsequently, we further explore the structural design details of the STAR-RIS system and finally present the design of an 8 $\times$ 4 STAR-RIS prototype, demonstrating its promising application prospects through performance evaluation results.

\subsection{Design of the Circuit}
\label{subsec:E_board design}

The signal processing circuit integrates three key functions: signal amplification, phase shifting, and power distribution. Signal amplification is achieved using the GALI-S66+ power amplifier chip from Mini-Circuits, which enhances the input signal strength. The phase shifting function is implemented using a 1-bit quantized phase control circuit, which consists of two MXD8625 SPDT switches from Maxscend and two phase delay lines between the switches. Power distribution is managed by a tunable power divider composed of two ${\lambda/8}$ coupled lines and two BB857 varactor diodes from Infineon. This design provides flexibility and tunability in controlling signal phase and power distribution, contributing to improved performance in RIS applications.

\begin{figure}[!htbp]
	\centering
	\includegraphics[width=0.3\textwidth]{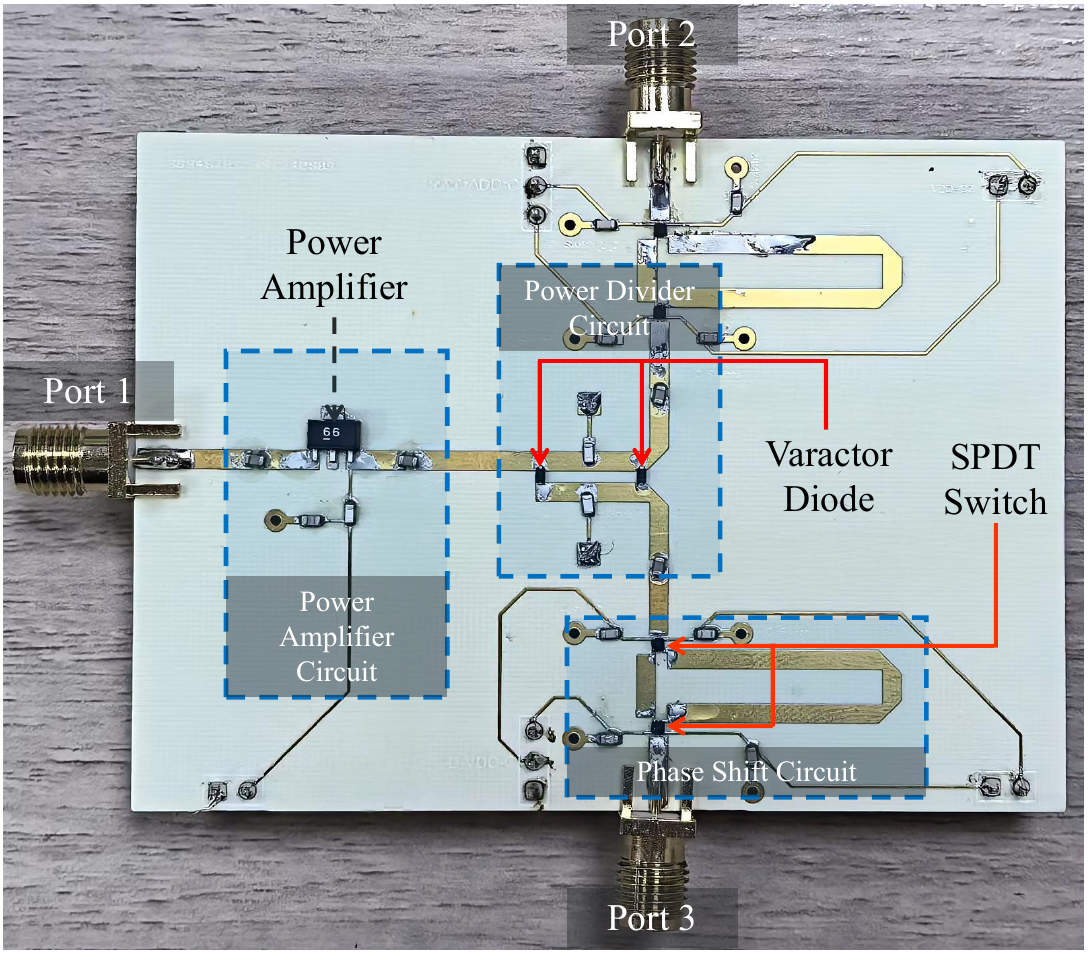}
	\caption{Photograph of the circuit evaluation board.}
	\label{fig:E-board}
\end{figure}

To validate the effectiveness of this circuit design, we developed an evaluation board, as shown in Fig.~\ref{fig:E-board}. The substrate is composed of Rogers 4350B material ($\varepsilon_r$ = 3.66, $\tan \delta$ = 0.0037) with a thickness of 0.762 mm. The input signal is first amplified using a GALI-S66+ power amplifier. Under the conditions of a 3.5 V supply voltage and 20mA operating current, this amplifier provides a gain of no less than 15 dB, thereby establishing a foundation for achieving high gain in subsequent active STAR-RIS designs while maintaining low power consumption. The evaluation board incorporates a BB857 varactor diode, whose capacitance varies between 0.45 pF and 7.2 pF, and is applied across both ends of a pair of one-eighth wavelength coupled lines. By adjusting the bias voltage, a dynamic power distribution of nearly 10 dB across two ports can be achieved, meeting the need for a system for precise signal distribution control. After passing through the power divider, the signal is routed into two independent 1-bit phase control circuits. Each phase control circuit consists of two MXD8625 SPDT switches. According to the datasheet, the MXD8625 switch exhibits a low insertion loss of approximately 0.2 dB within the target frequency band. Switches A and B are symmetrically arranged, with the control signal of switch A ($VCC_A$) and that of switch B ($VCC_B$) being inversely related. $VCC_A$ operates at a high level of 2.8 V and a low level of 0 V. When $VCC_A$ is high, and $VCC_B$ is low, the output phase is 0°; conversely, when $VCC_A$ is low, and $VCC_B$ is high, the output phase is 180°. Since the two phase control circuits are independent, the system can achieve independent phase control of both transmitted and reflected signals in the STAR-RIS, thus enhancing the ability of the system to manage multipath signals.

\begin{figure}[!htbp]
    \centering 
    \subfigure[]{
    \includegraphics[width=0.23\textwidth]{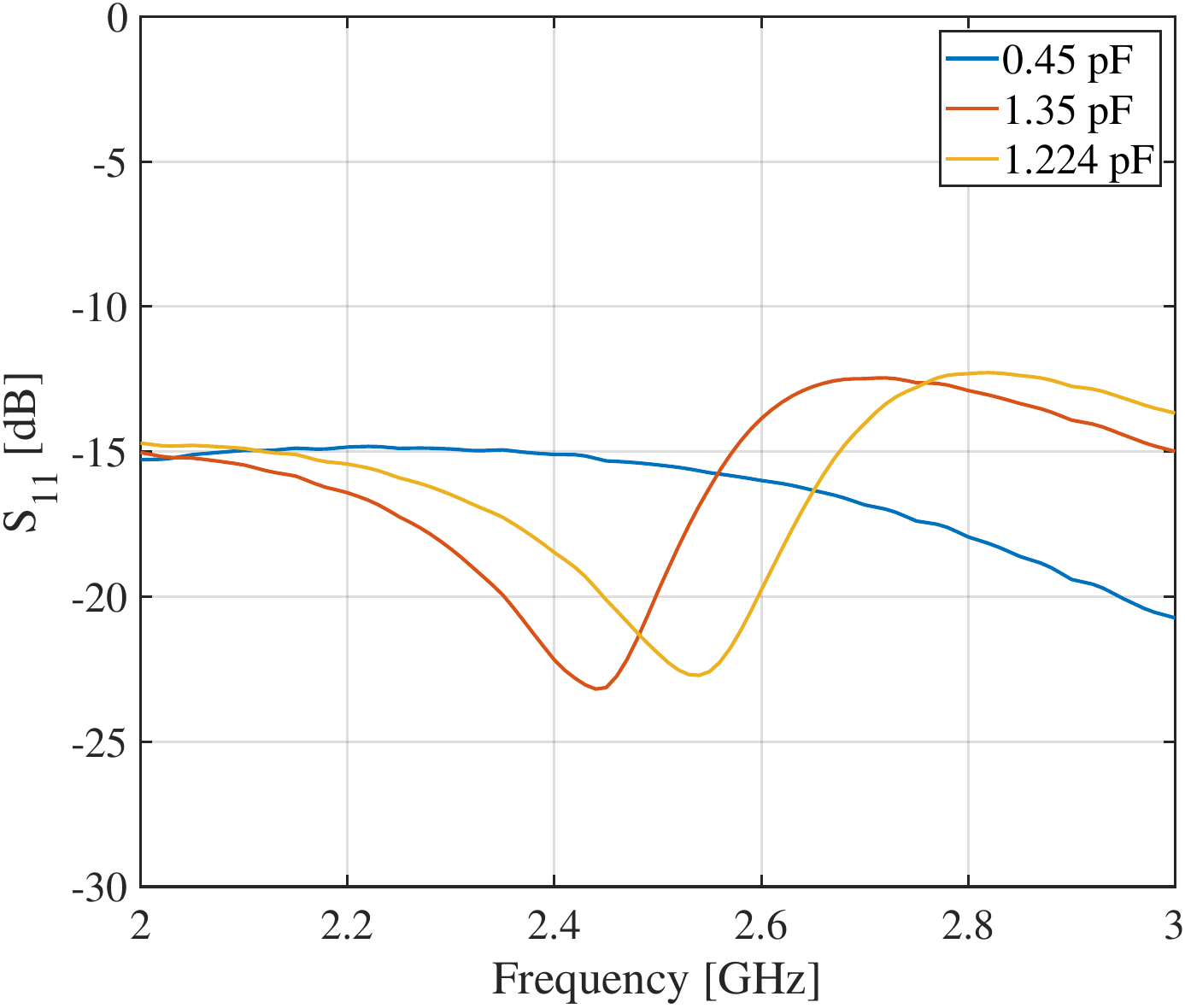}\label{fig:E_board_S_S_11}}
     \subfigure[]{
    \includegraphics[width=0.23\textwidth]{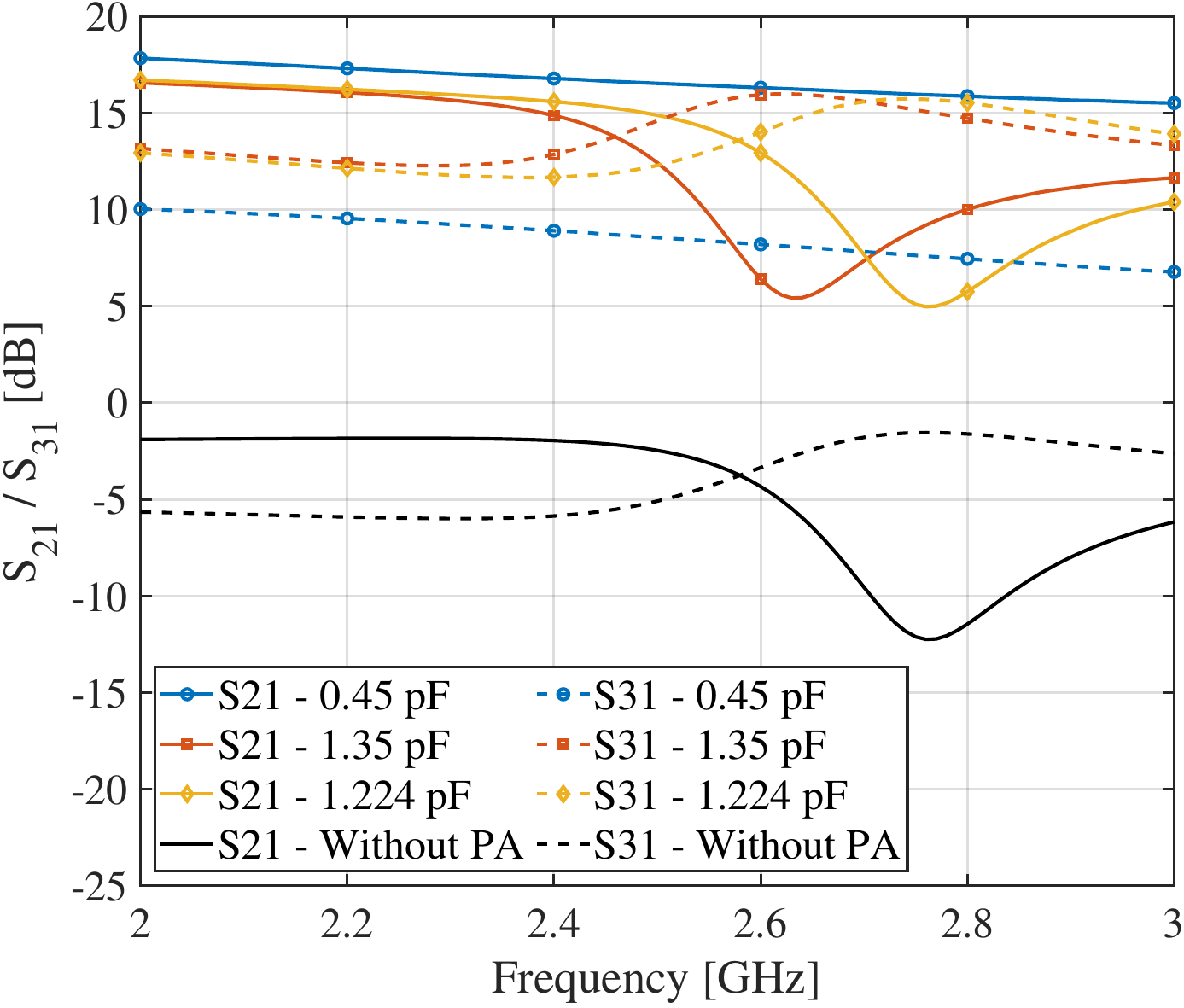}\label{fig:E_board_S_S_21_31}}
    \subfigure[]{
    \includegraphics[width=0.23\textwidth]{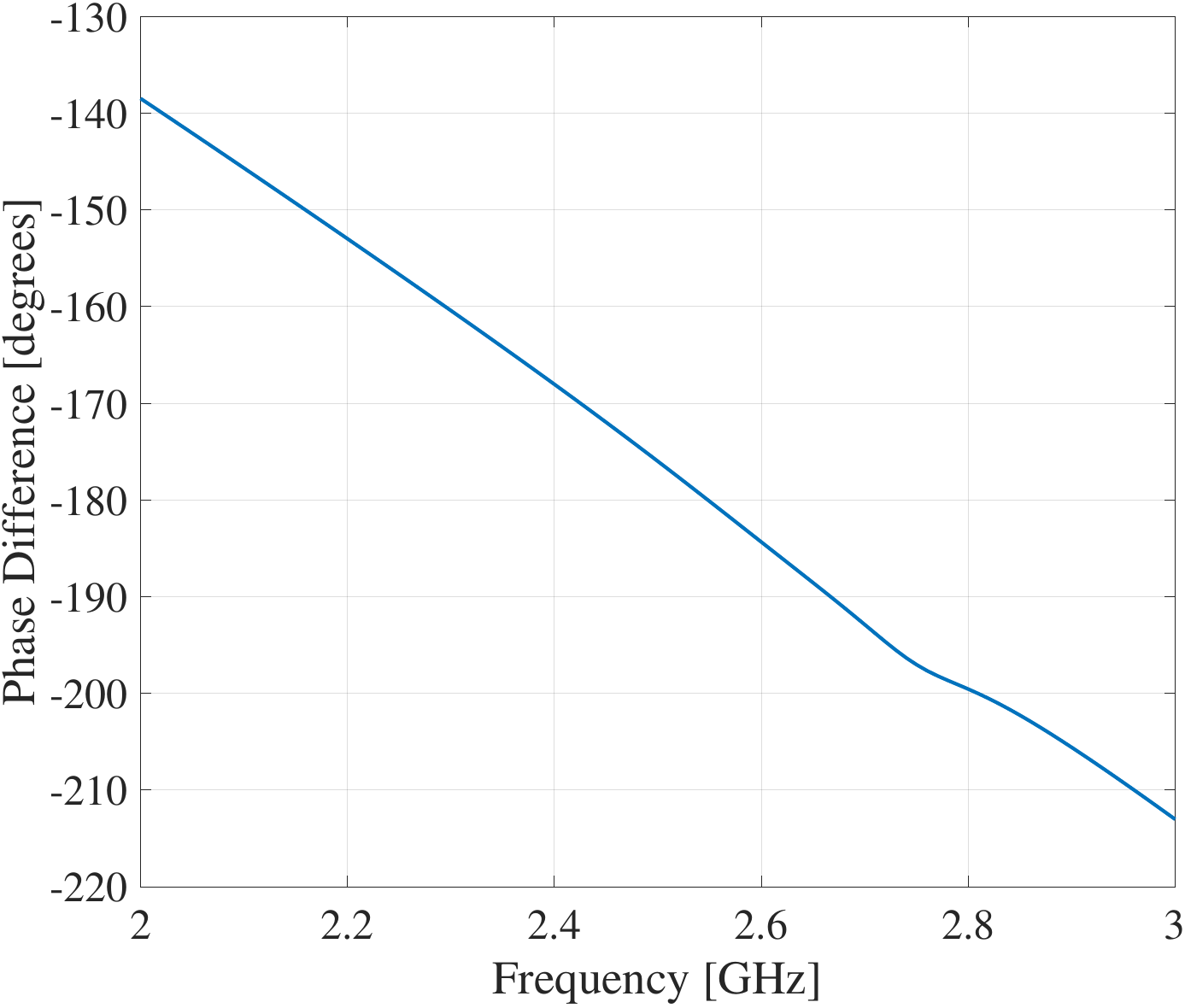}\label{fig:E_board_S_phase}}
        \subfigure[]{
    \includegraphics[width=0.23\textwidth]{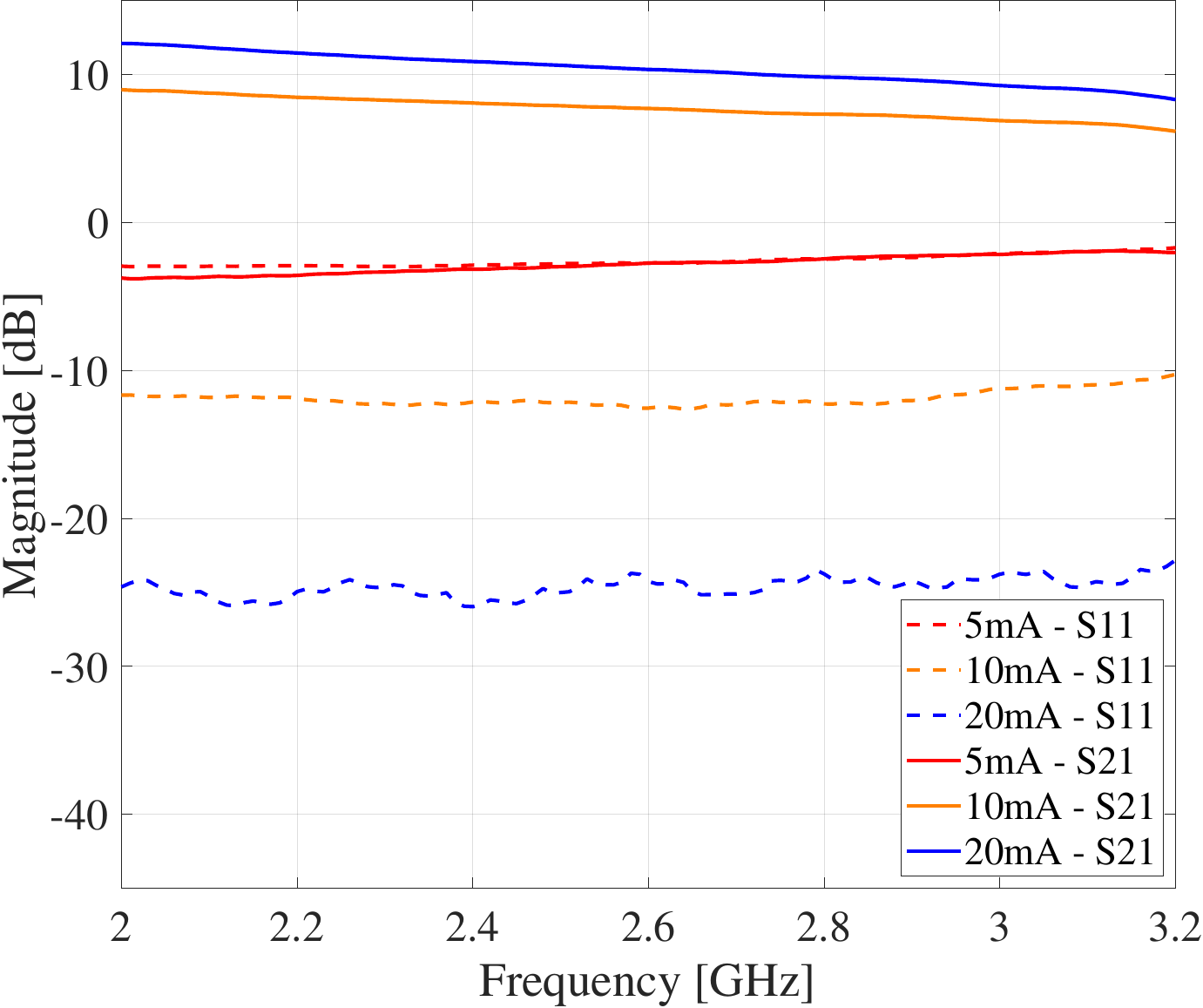}\label{fig:E-BOARD_PA_stage}}
     \subfigure[]{
    \includegraphics[width=0.23\textwidth]{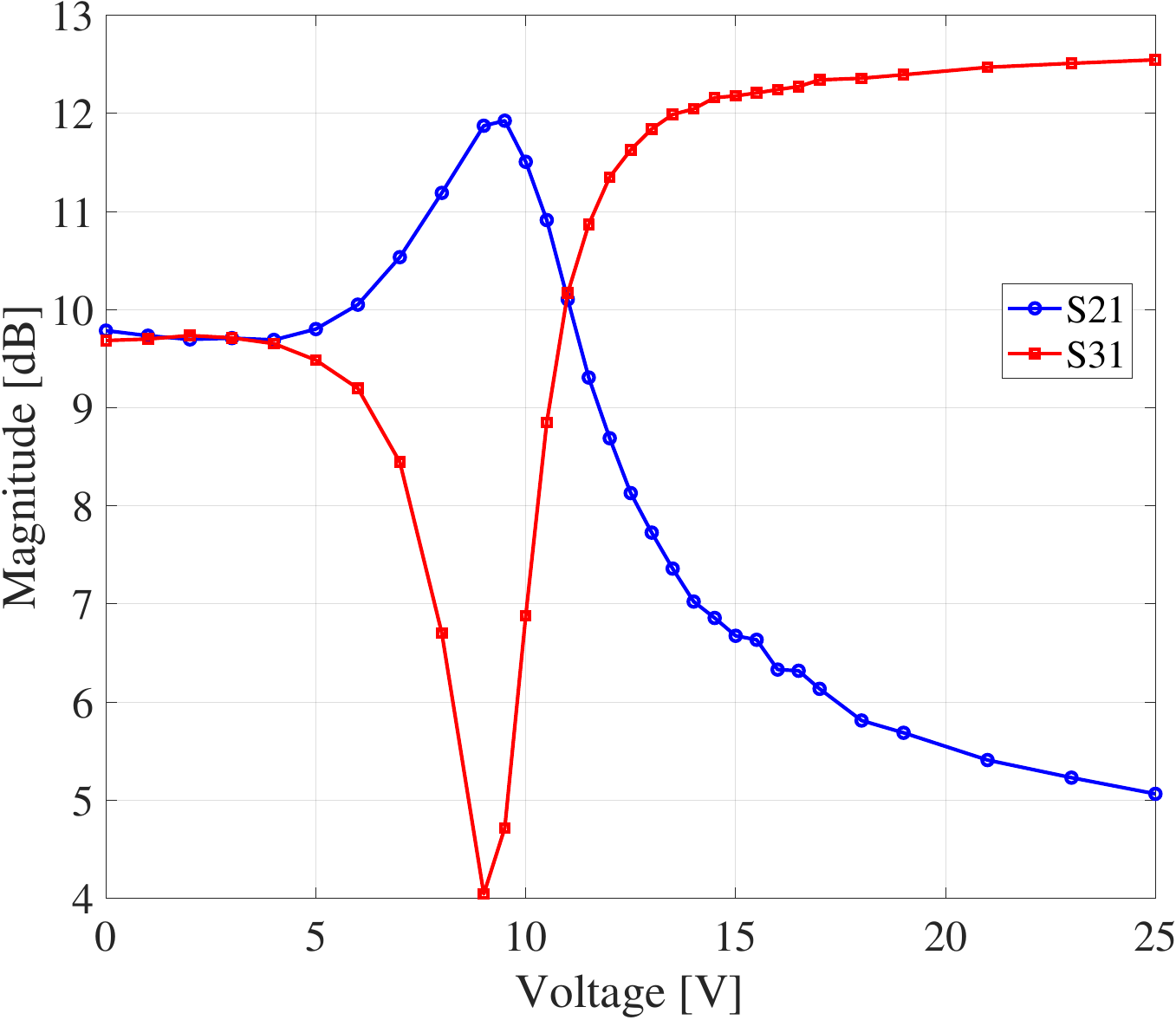}\label{fig:E-BOARD_DIODE}}
    \subfigure[]{
    \includegraphics[width=0.23\textwidth]{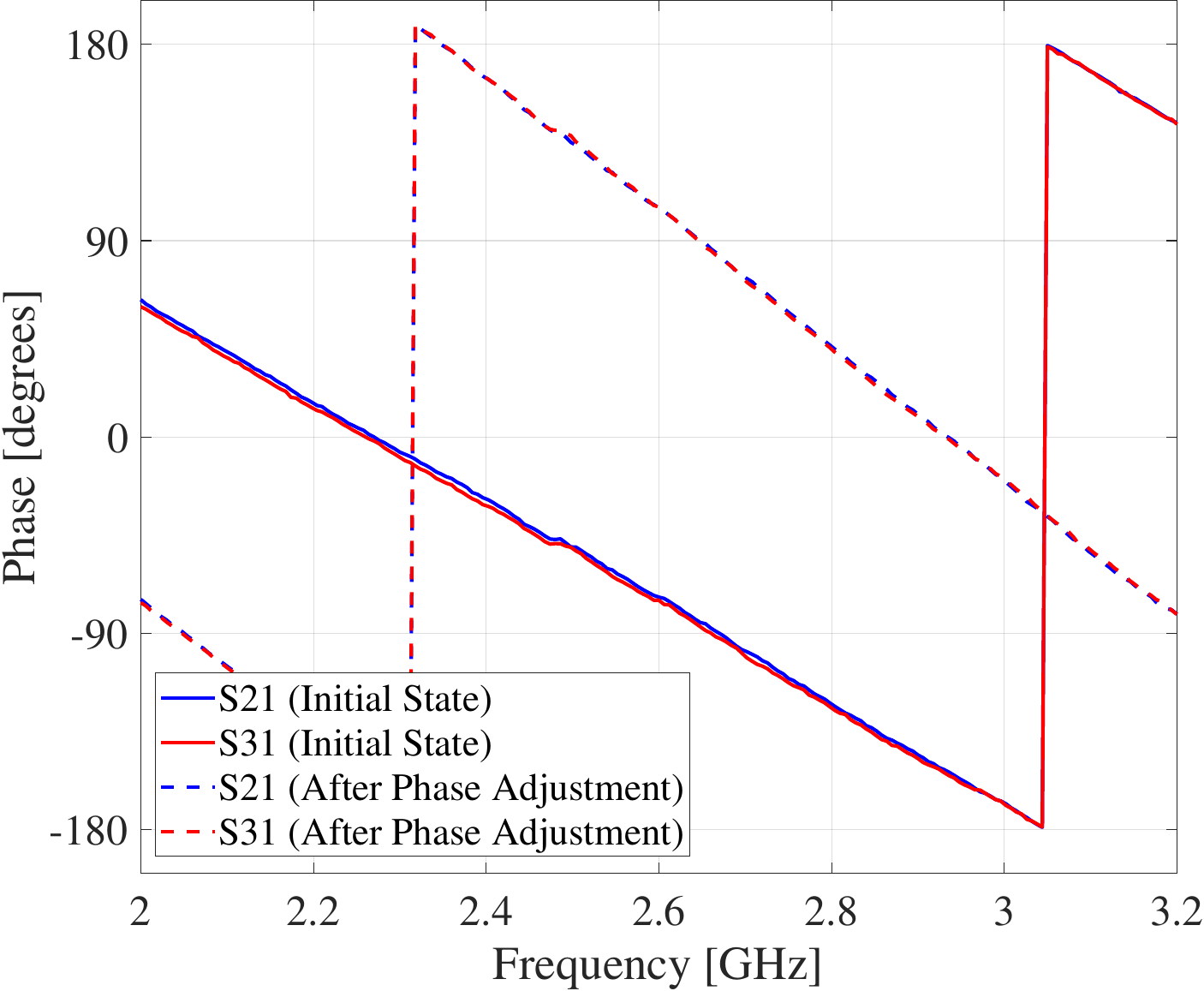}\label{fig:E-BOARD_Phase}}
    \caption{Simulation and measurement results of the evaluation board:
(a) Simulated reflection coefficient ($S_{11}$) performance.
(b) Simulated adjustable power split ratio between ports ($S_{21}$ vs. $S_{31}$).
(c) Simulated phase difference control for 0°/180° phase shift.
(d) Measured active gain variation of the power amplifier under different operating currents.
(e) Measured output power trends of ports 2 and 3 with respect to voltage, demonstrating voltage-dependent power distribution.
(f) Measured phase control performance validation.}

    \label{fig:Results of E_board}
\end{figure}

The simulation results, shown in Fig.~\ref{fig:Results of E_board}, provide a detailed depiction of the circuit behavior under various operating conditions. Specifically, Fig.~\ref{fig:E_board_S_S_11} demonstrates that the reflection coefficient ($S_{11}$) remains below -10 dB across the desired frequency band under different power distribution conditions, indicating that the circuit effectively maintains good impedance matching over the entire operating frequency range, thus maximizing signal transmission efficiency.

Regarding power distribution and capacitance control, the simulation results show that when the capacitance of the BB857 varactor diode is 1.224 pF, the circuit achieves equal power output at 2.6 GHz, as shown in Fig.~\ref{fig:E_board_S_S_21_31}. In the absence of the power amplifier (PA) gain, the insertion losses $S_{21}$ and $S_{31}$ are nearly identical, both around -4 dB, indicating low insertion loss in the circuit. Upon introducing the power amplifier, the circuit continues to deliver equal power output at the same capacitance value while the amplifier adds approximately 15 dB of additional gain, further enhancing the overall system performance. This result verifies the effective operation of the power amplifier, which is also corroborated by the measured data. As shown in Fig.~\ref{fig:E-BOARD_PA_stage}, at a power amplifier operating current of 20 mA, the circuit achieves maximum power output, with $S_{21}$ around 10 dB. As the operating current of the amplifier varies, the active gain provided by the system shows notable adjustability, further proving the tunability of the design.

Further analysis of the effect of varying the varactor diode capacitance on the output ports was conducted by adjusting the capacitance of the varactor diode in the simulation. The results, as shown in Fig.~\ref{fig:E_board_S_S_21_31}, reveal that when the capacitance of the BB857 varactor diode decreases to a minimum value of 0.45 pF, $S_{21}$ is significantly larger than $S_{31}$, with a difference of nearly 10 dB. Conversely, when the capacitance increases to 1.35 pF, $S_{31}$ becomes significantly greater than $S_{21}$ at 2.6 GHz. These findings demonstrate that the capacitance of the varactor diode has a substantial impact on the power split ratio and can be precisely adjusted to control the power distribution between the output ports.

The measured results further validate these simulation trends. At 2.6 GHz, we measured the magnitude of $S_{21}$ and $S_{31}$ as a function of the bias voltage applied across the varactor diode. As shown in Fig.~\ref{fig:E-BOARD_DIODE}, the measured results are in good agreement with the simulation, indicating that when the bias voltage is set to 11 V, equal power output is achieved at both output ports. Additionally, as the voltage varies, the magnitude of the output at both ports exhibits a tunable range of nearly 8 dB, with the adjustment being bidirectional, which further validates the efficiency and tunability of the circuit in the power distribution control.

Regarding the phase control, both simulation and measurement results confirm that by selecting different signal paths via switches, the circuit can achieve a phase difference of approximately 180 degrees between the two states at 2.6 GHz, demonstrating the precise phase control capability of the circuit (Fig.~\ref{fig:E_board_S_phase}). The measured results in Fig.~\ref{fig:E-BOARD_Phase} further validate the effectiveness of the design in controlling the phase between the two ports, showing that the system can precisely adjust the phase of the signals, thereby enhancing its ability to handle multipath signals and improve overall system performance.

\subsection{Design Details of the STAR-RIS Unit}
\label{subsec:Unit design}

\begin{figure}[!htbp]
    \centering 
    \subfigure[]{
    \includegraphics[width=0.23\textwidth]{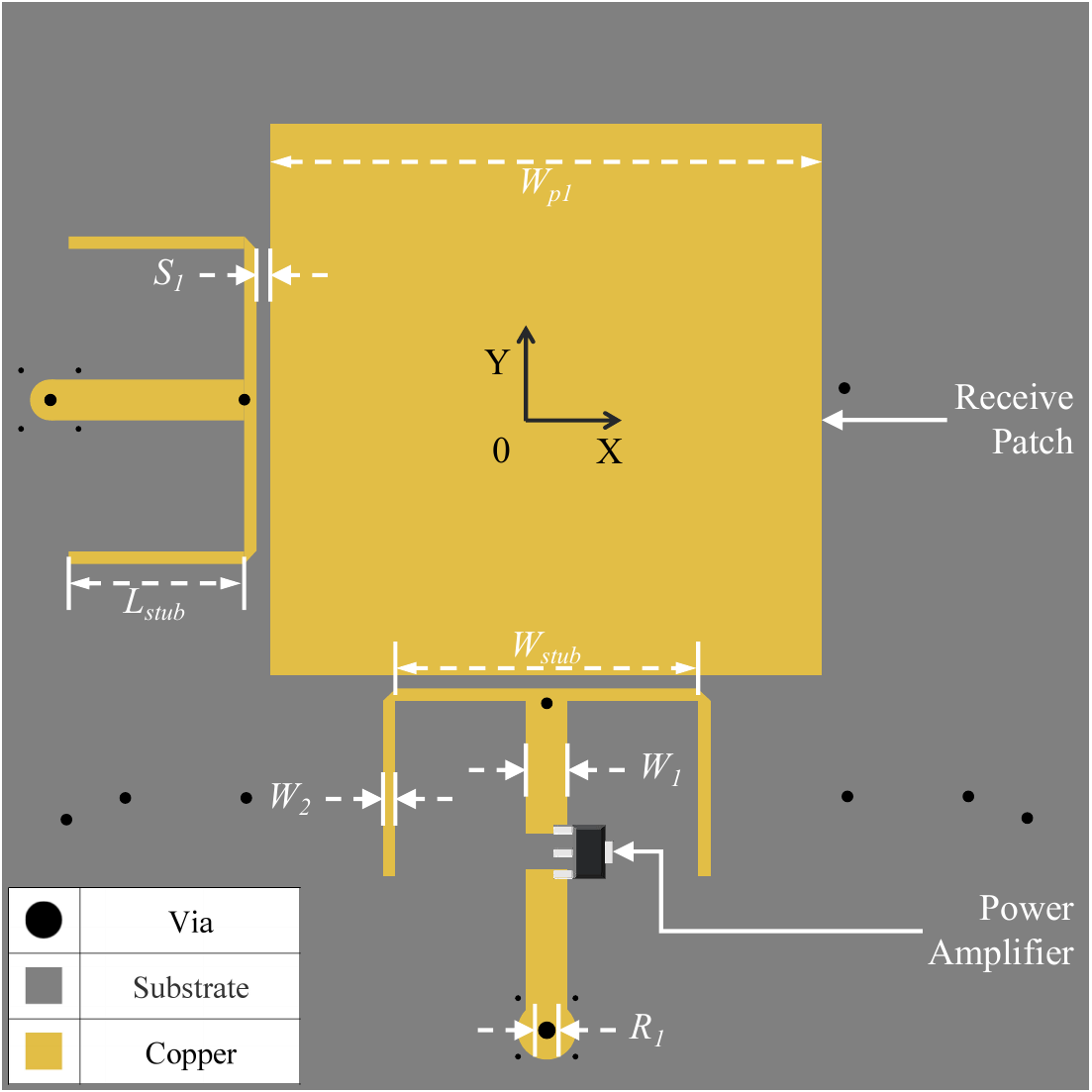}\label{fig:unit top}}
     \subfigure[]{
    \includegraphics[width=0.23\textwidth]{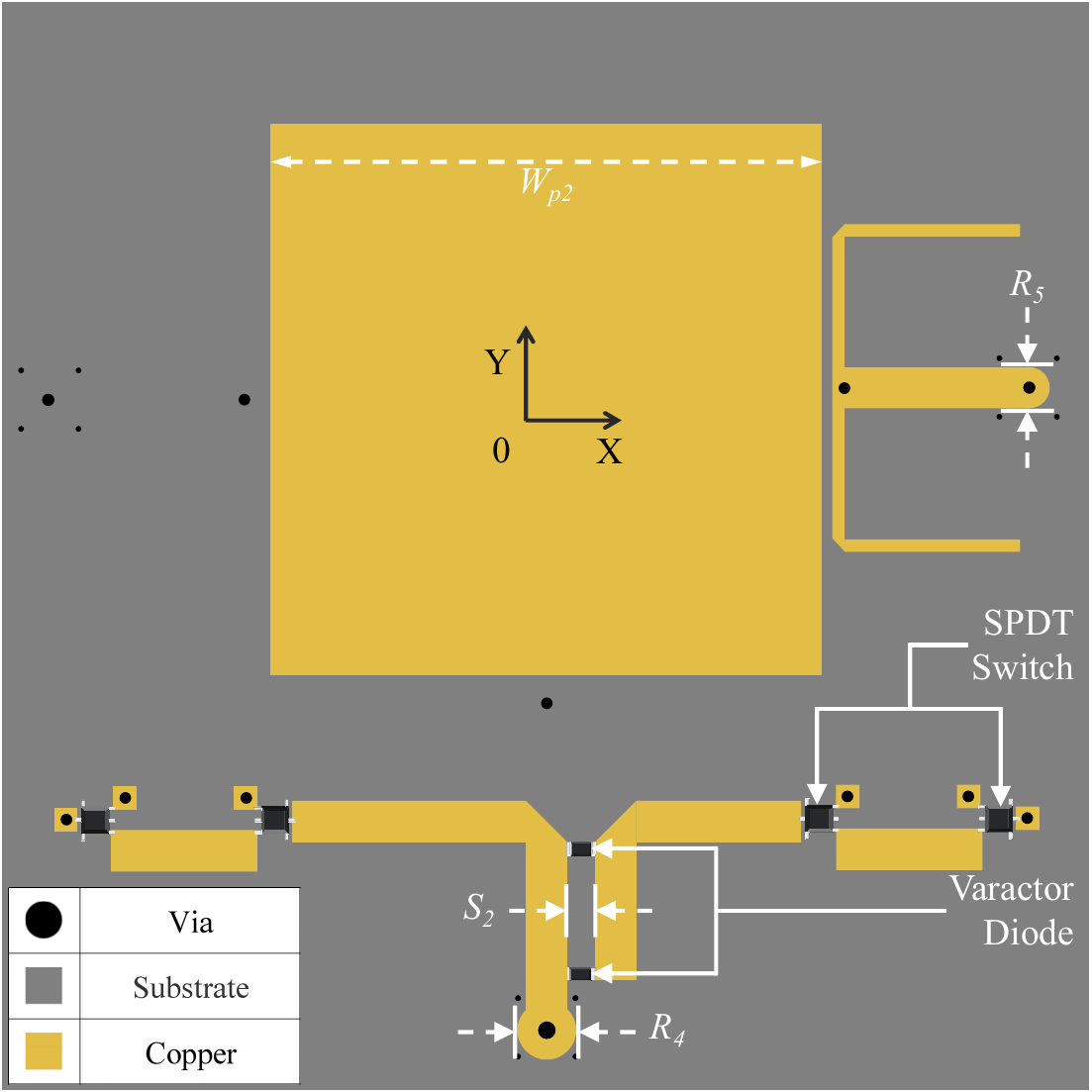}\label{fig:unit bottom}}
    \subfigure[]{
    \includegraphics[width=0.23\textwidth]{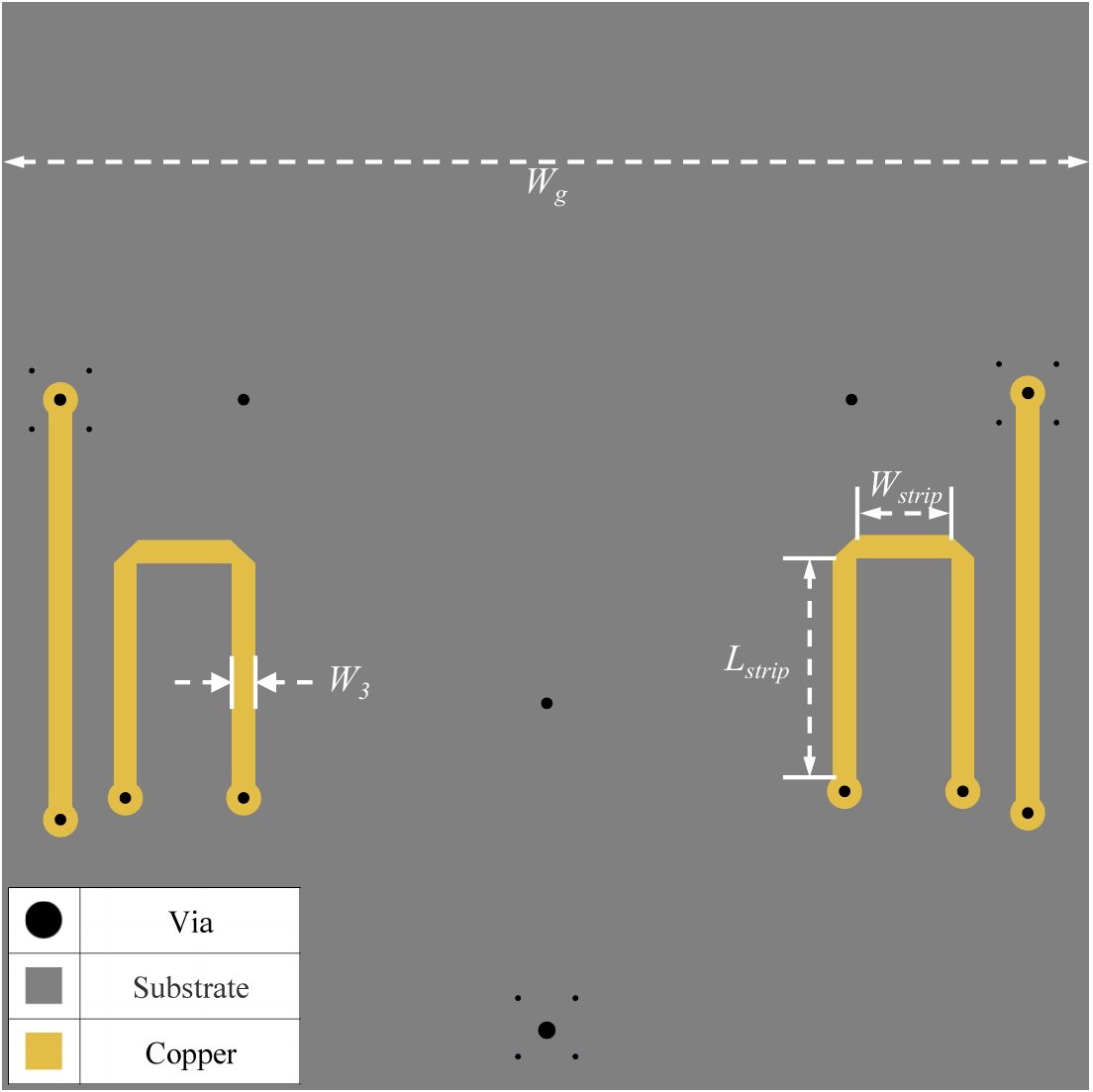}\label{fig:unit mid}}
     \subfigure[]{
    \includegraphics[width=0.23\textwidth]{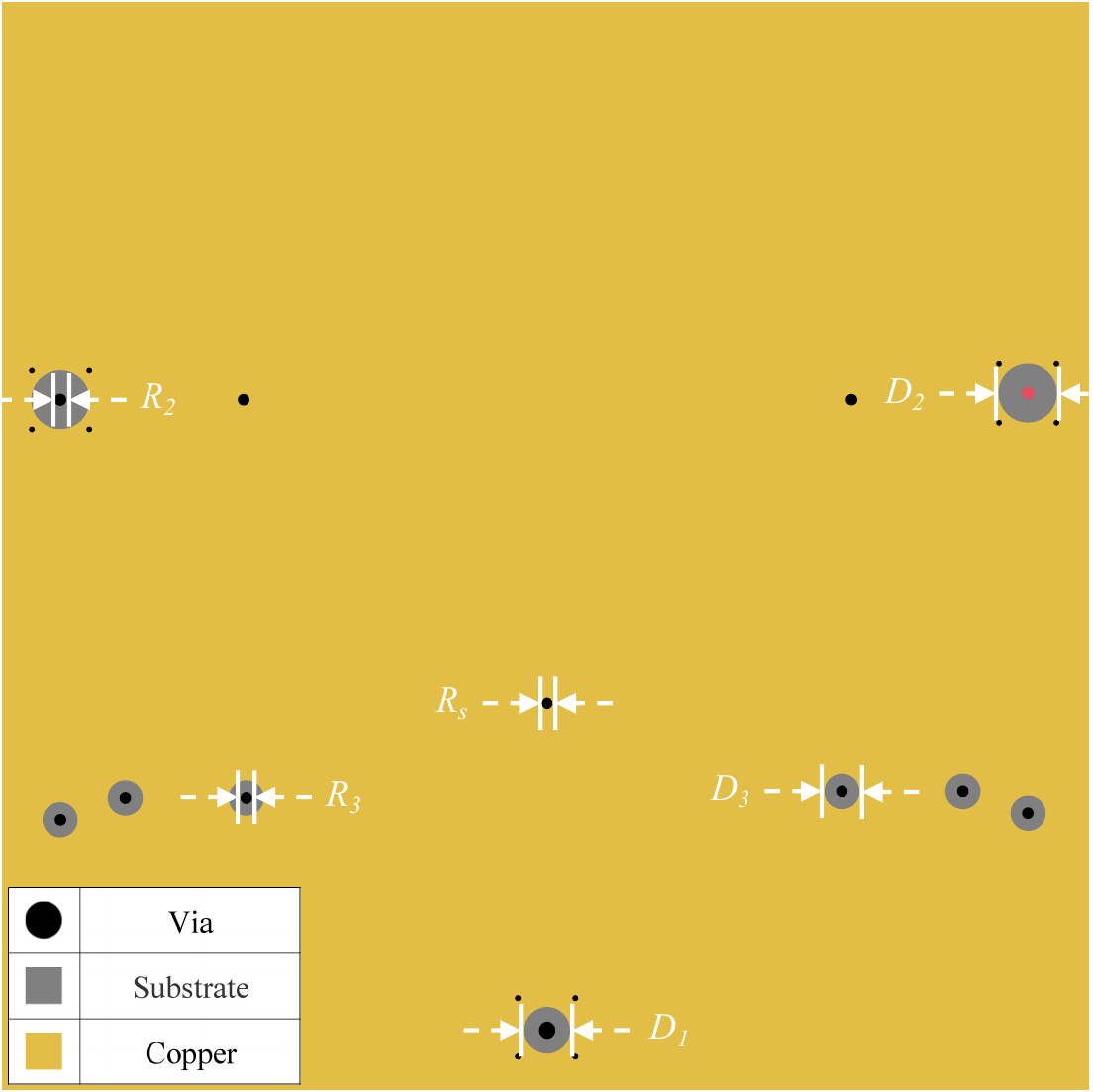}\label{fig:unit GND}}\\
     \subfigure[]{
    \includegraphics[width=0.23\textwidth]{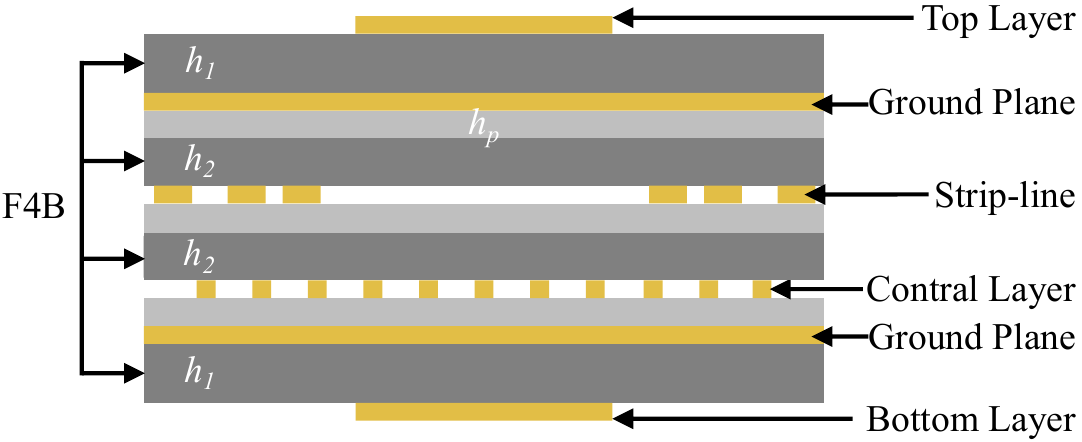}\label{fig:unit layer}}
    \caption{Schematic diagram of STAR-RIS unit: (a) Top layer. (b) Bottom layer. (c) Strip-line layer. (d) Ground plane. (e) Lateral-view of the STAR-RIS unit.}
    \label{fig:Schematic diagram of STAR-RIS unit}
\end{figure}

Based on the signal processing circuit design proposed in Sec. \ref{subsec:E_board design}, we further present the design of the active STAR-RIS unit. The detailed structure of the unit is shown in Fig.~\ref{fig:Schematic diagram of STAR-RIS unit}. The overall structure consists of two F4B substrates with a thickness of 1 mm, two F4B substrates with a thickness of 0.508 mm, and multiple prepreg layers placed between the substrates to achieve multilayer bonding. The dielectric constant of the substrates ($\varepsilon_r$) is 3.5, with a loss tangent ($\tan \delta$) of 0.002, while the prepreg layers have the same dielectric constant of 3.5 and a loss tangent of 0.003. The multilayer structure is depicted in Fig.~\ref{fig:unit layer} for further clarification. The unit cell dimensions of 58 mm $\times$ 58 mm indicate that the periodicity of the array corresponds to half of the free-space wavelength.

\begin{table}[!htbp]
\centering
\renewcommand{\arraystretch}{1.3}
\caption{The Structural Parameters of The STAR-RIS Unit.}
\setlength{\tabcolsep}{1mm}{
\label{tab:Structural Parameters of RIS Unit}
\centering
\begin{tabular}{|c|c||c|c||c|c|}
\hline\hline
\textbf{Parameter} & \textbf{Size} & \textbf{Parameter} & \textbf{Size} & \textbf{Parameter} & \textbf{Size} \\\hline
$W_\text{p1}$      & 29.8 mm       & $W_\text{p2}$      & 30 mm         & $W_\text{stub}$    & 18 mm         \\\hline
$L_\text{stub}$    & 9.1 mm        & $W_\text{strip}$   & 6.3 mm        & $L_\text{strip}$   & 13.2 mm       \\\hline
$W_\text{g}$       & 58 mm         & $W_\text{1}$       & 2.2 mm        & $W_\text{2}$       & 0.3 mm        \\\hline
$W_\text{3}$       & 0.8 mm        & $S_\text{1}$       & 1.15 mm       & $S_\text{2}$       & 1.5 mm        \\\hline
$R_\text{1}$       & 0.5 mm        & $R_\text{2}$       & 0.6 mm        & $R_\text{3}$       & 0.4 mm        \\\hline
$R_\text{4}$       & 3 mm          & $R_\text{5}$       & 2.2 mm        & $R_\text{s}$       & 0.4 mm        \\\hline
$D_\text{1}$       & 2.5 mm        & $D_\text{2}$       & 3 mm          & $D_\text{3}$       & 2 mm          \\\hline
$h_\text{1}$       & 1 mm          & $h_\text{2}$       & 0.508 mm      & $h_\text{p}$       & 0.2 mm        \\\hline
\end{tabular}}

\end{table}

The top layer structure of the STAR-RIS unit is illustrated in Fig.~\ref{fig:unit top}; this layer primarily serves to receive the incident electromagnetic waves, amplify the incoming signals, and radiate the reflected signals. The top layer consists of a square patch, a pair of orthogonally oriented T-shaped coupling feeds, a power amplification circuit, and the associated power supply circuitry. The 28.8 mm $\times$ 28.8 mm patch is designed to receive incident electromagnetic wave signals and also functions as the radiating antenna for reflecting these signals. To facilitate out-of-band filtering capabilities, T-shaped coupling feeds are utilized. Each feed incorporates two symmetrically placed L-shaped quarter-wavelength short-circuit resonators coupled with a 50~$\Omega$ microstrip line. These L-shaped resonators are interconnected to the ground plane through a shared metallic short-circuit via. This configuration not only augments the resonant points, thereby extending the bandwidth but also efficiently suppresses harmonic interference, thereby enabling effective out-of-band filtering.
 The Y-axis feed is responsible for transmitting the incident electromagnetic signals and amplifying them using a power amplifier chip also aligned along the Y-axis. Conversely, the X-axis feed transmits the reflected signals, which are then radiated by the square patch. Metal vias are incorporated at the ends of both T-shaped coupling feeds to ensure electrical connectivity between the top and bottom layers.

The underlying structure of the STAR-RIS unit is depicted in Fig.~\ref{fig:unit bottom}. This layer primarily functions to distribute the power of the amplified incident signal, independently control the phase of reflected and transmitted signals, and radiate the transmitted signal. Similar to the first layer, the bottom layer also employs a square radiation patch to radiate the transmitted signal. To excite the patch and enable the transmission of the signal, a T-shaped coupling feed, identical in structure to that on the top layer for reflecting signal excitation, is symmetrically placed on the bottom layer. It is important to note that the dimensions of the bottom layer differ slightly from the top layer in terms of parameters. Unlike the top layer, the bottom layer incorporates a pair of 50 $\Omega$ quarter-wavelength coupling lines and two varactor diodes to dynamically control the power distribution between the reflected and transmitted signals, consistent with the design outlined in Sec. \ref{subsec:E_board design}. Additionally, four SPDT switches are symmetrically arranged in two pairs on the bottom layer. These switches, connected through microstrip lines on the bottom layer and strip lines on the middle layer (as shown in Fig.~\ref{fig:unit mid}), form two signal paths with a 180° phase difference. Through the switch control circuit located on the bottom layer, phase switching between 0° and 180° is achieved, enabling independent 1-bit phase control of the reflected and transmitted signals.

In addition to the structure as mentioned above, the STAR-RIS unit also includes two strip lines in the middle layer, which are independently connected to the output of the phase control circuit and the T-shaped feedlines for the transmitted and reflected signals (as shown in Fig.~\ref{fig:unit mid}). Two identical ground planes are placed on the lower surface of the first dielectric layer and the upper surface of the fourth dielectric layer (as shown in Fig.~\ref{fig:unit GND}). Between these two ground planes, in addition to the strip lines, a control circuit layer is deployed to manage the power supply and control signals for all components within the unit. To further suppress signal crosstalk and parasitic effects, several isolation vias are strategically placed to provide electromagnetic shielding, mitigating the crosstalk and parasitic inductance that may arise from through-hole vias (as shown in Fig.~\ref{fig:Schematic diagram of STAR-RIS unit}). These design elements enhance signal integrity and system stability, ensuring the efficient operation of the STAR-RIS unit.

\subsection{STAR-RIS Unit Simulation Results}
\label{subsec:Unit Simulation}

A pair of Floquet ports is placed approximately a quarter wavelength away from the top and bottom surfaces of the unit to enable plane wave incidence and reception. Additionally, master-slave boundary conditions are applied around the unit to simulate the array environment accurately. 

\begin{figure}[!htbp]
    \centering 
    \subfigure[]{
    \includegraphics[width=0.23\textwidth]{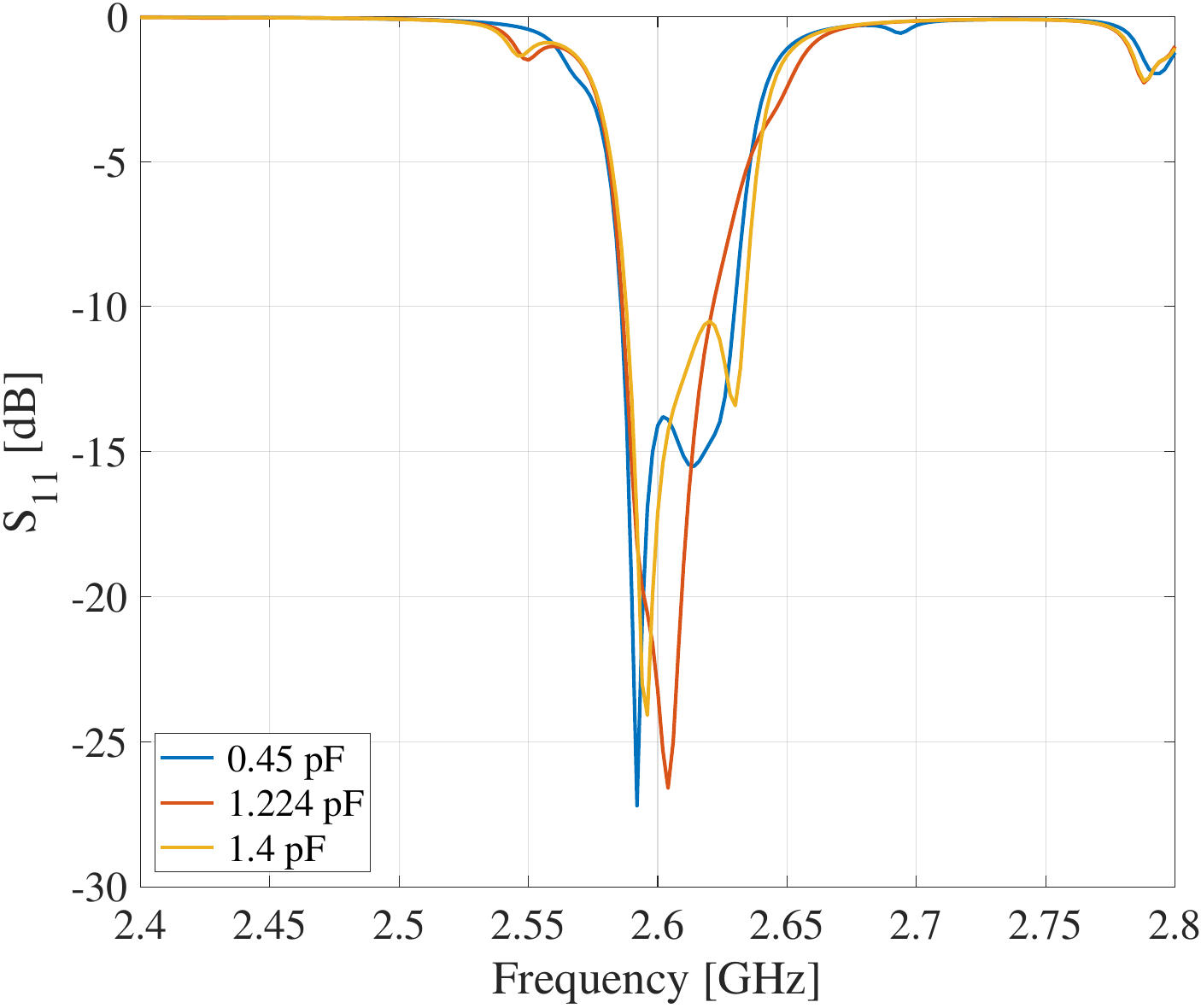}\label{fig:Unit_S11}}
     \subfigure[]{
    \includegraphics[width=0.23\textwidth]{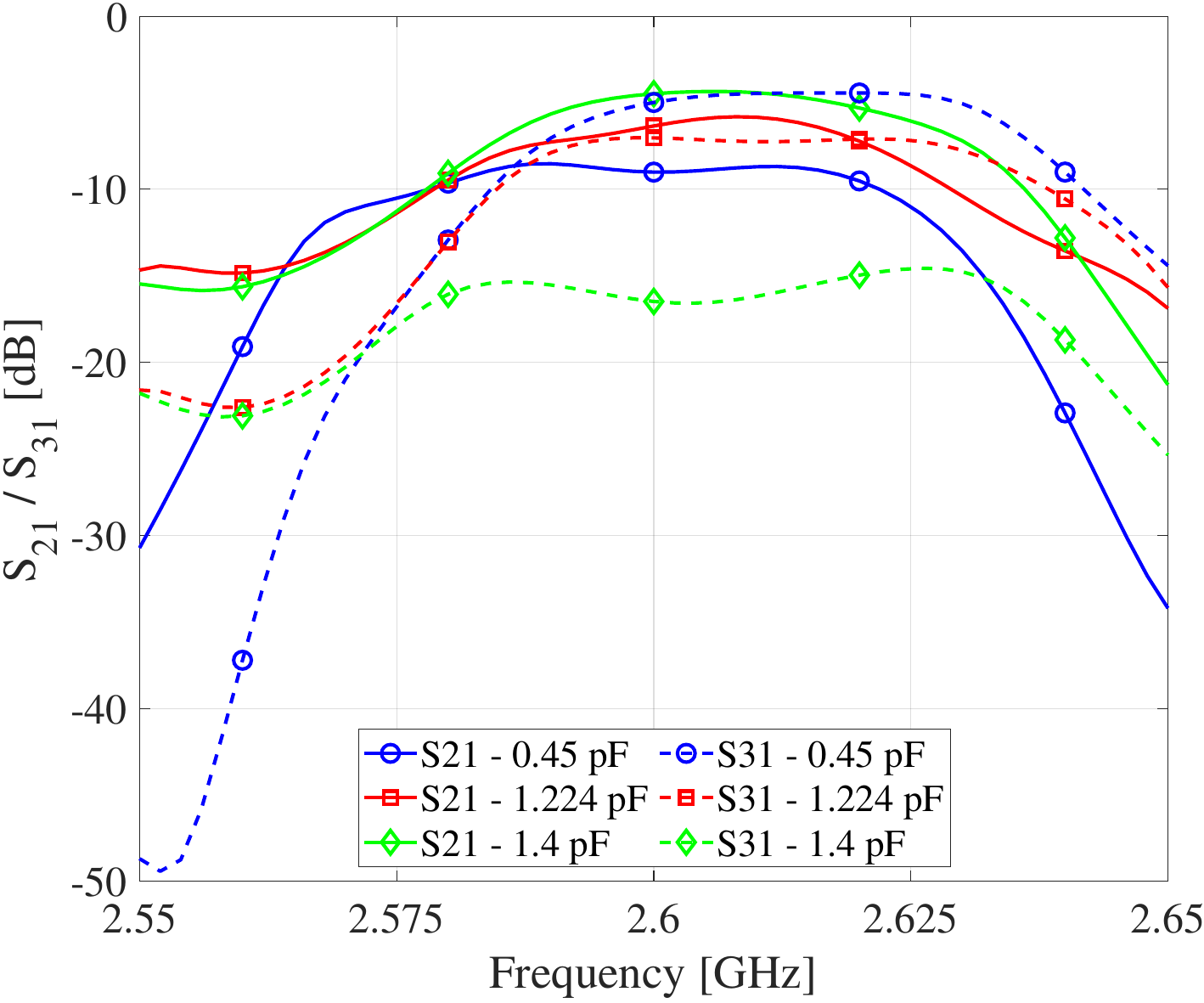}\label{fig:Unit_S21_31}}
    \subfigure[]{
    \includegraphics[width=0.23\textwidth]{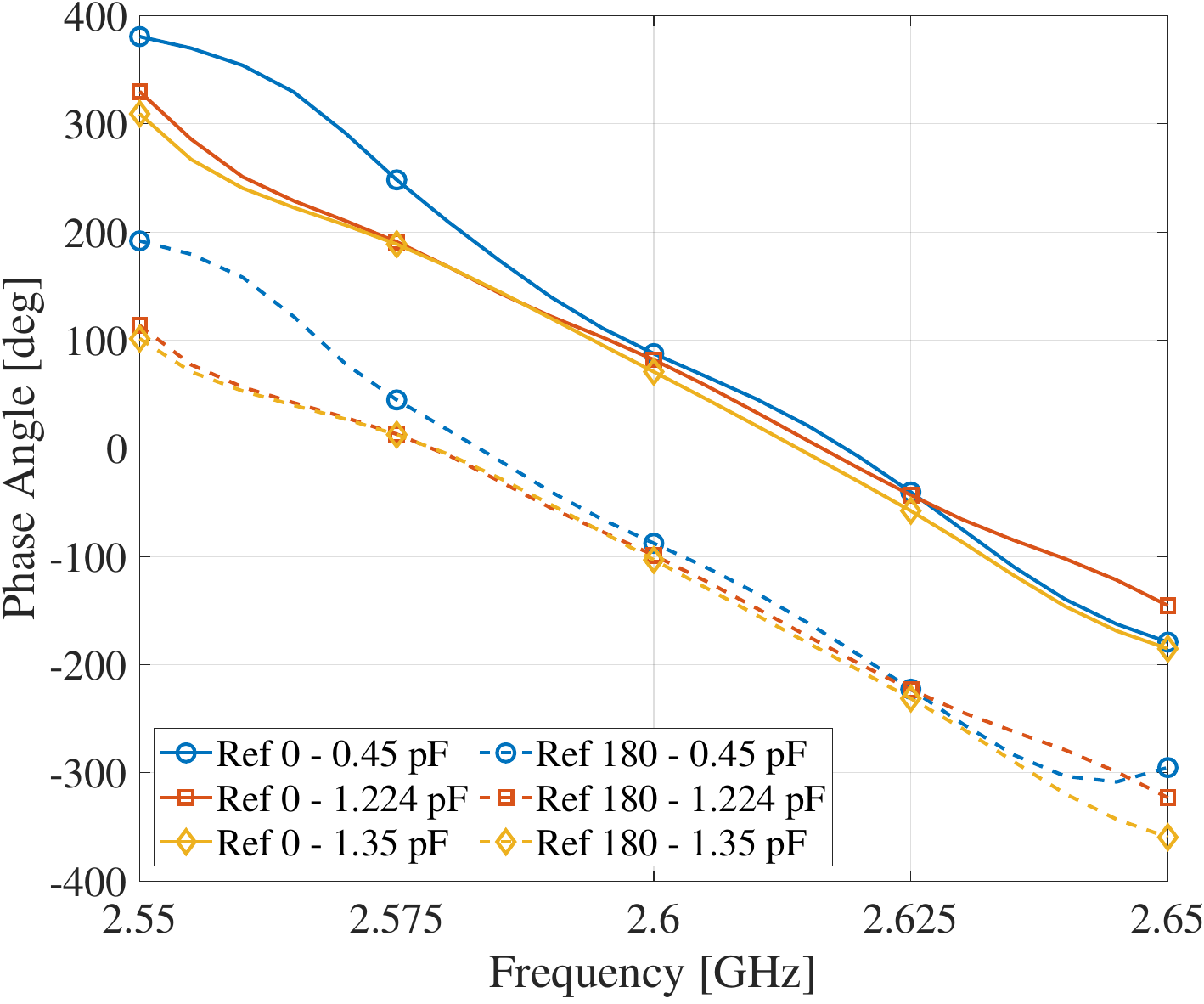}\label{fig:Ref_phase}}
    \subfigure[]{
    \includegraphics[width=0.23\textwidth]{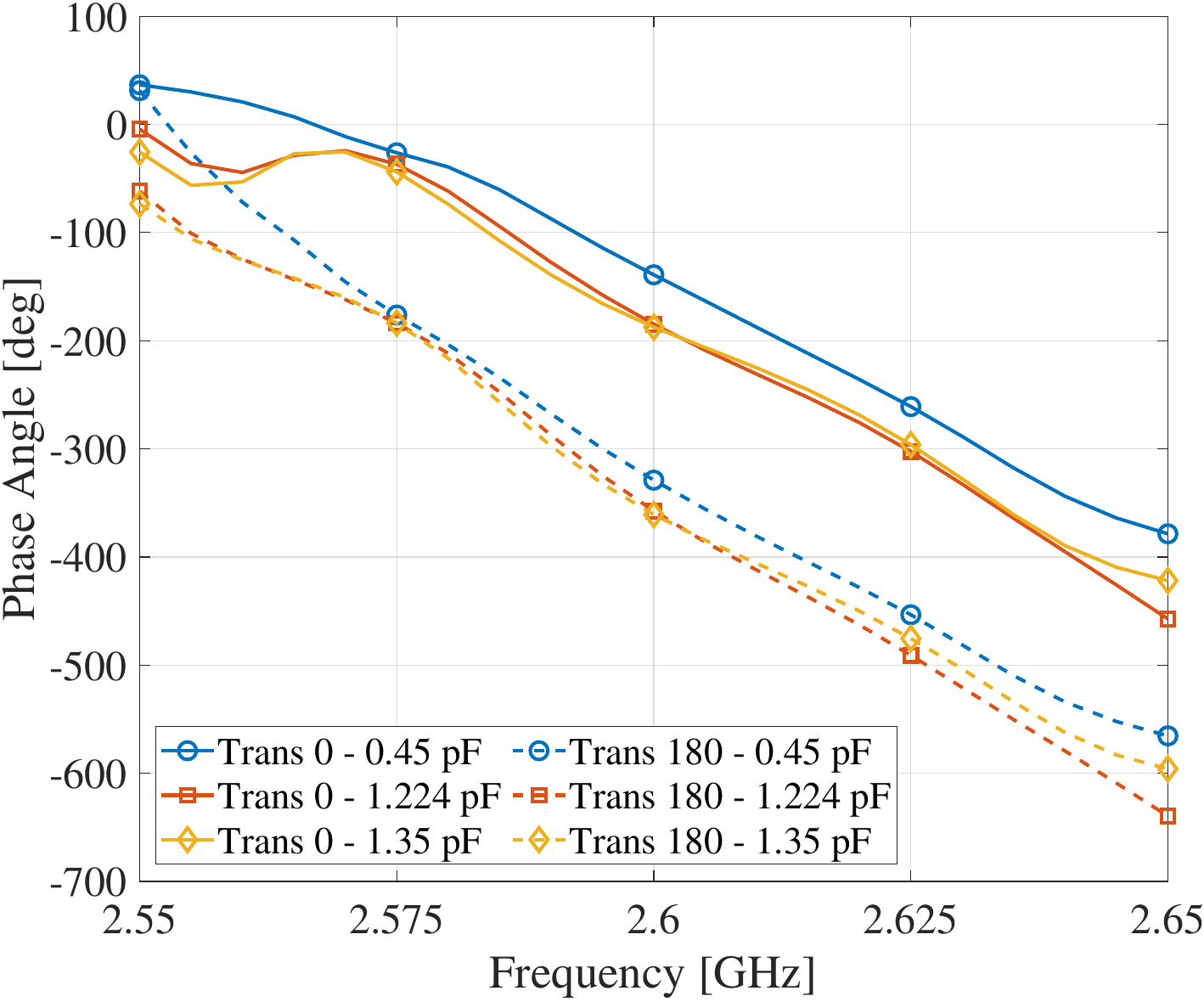}\label{fig:Trans_phase}}
     \subfigure[]{
    \includegraphics[width=0.23\textwidth]{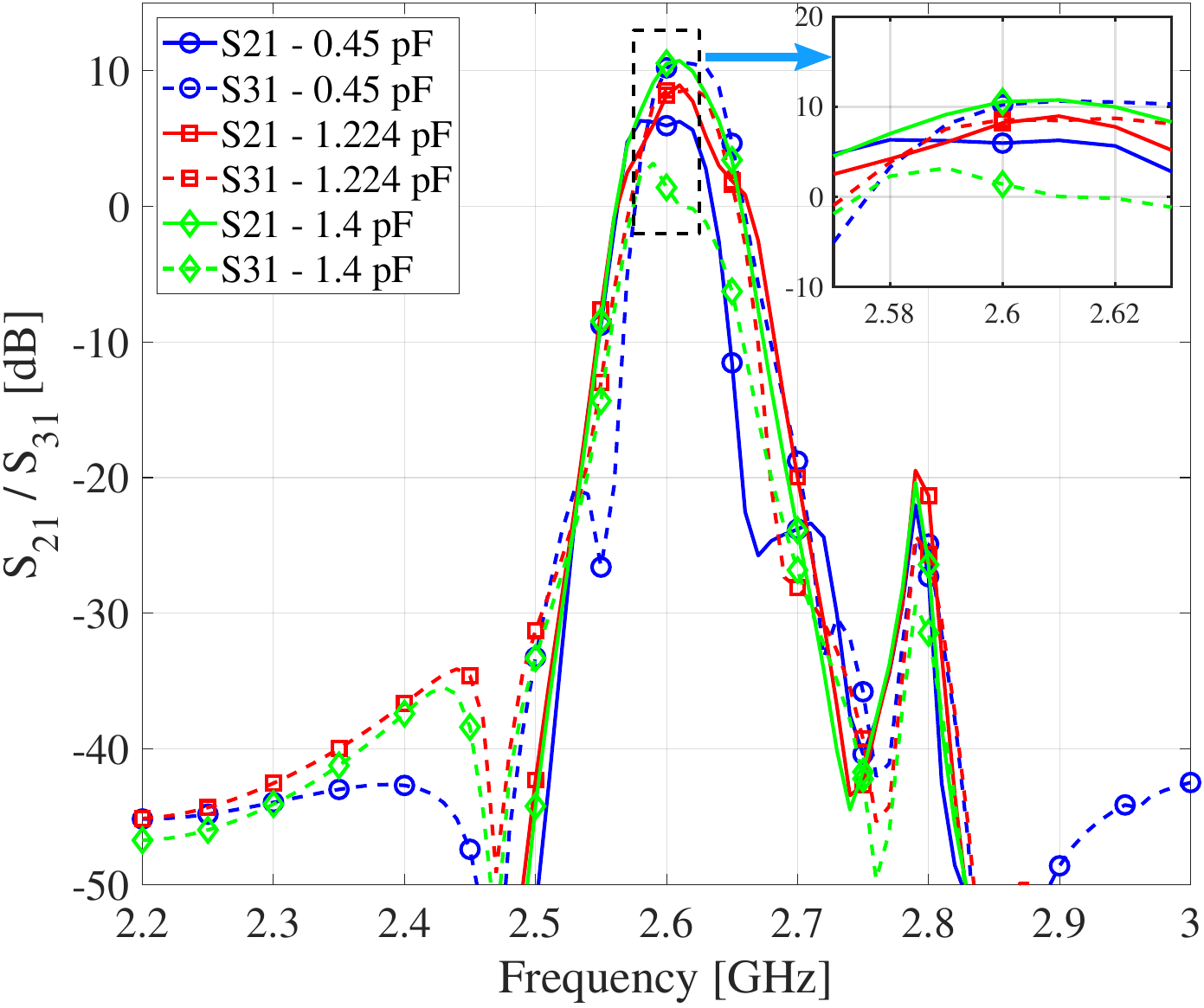}\label{fig:PA_S21_31}}
     \subfigure[]{
    \includegraphics[width=0.23\textwidth]{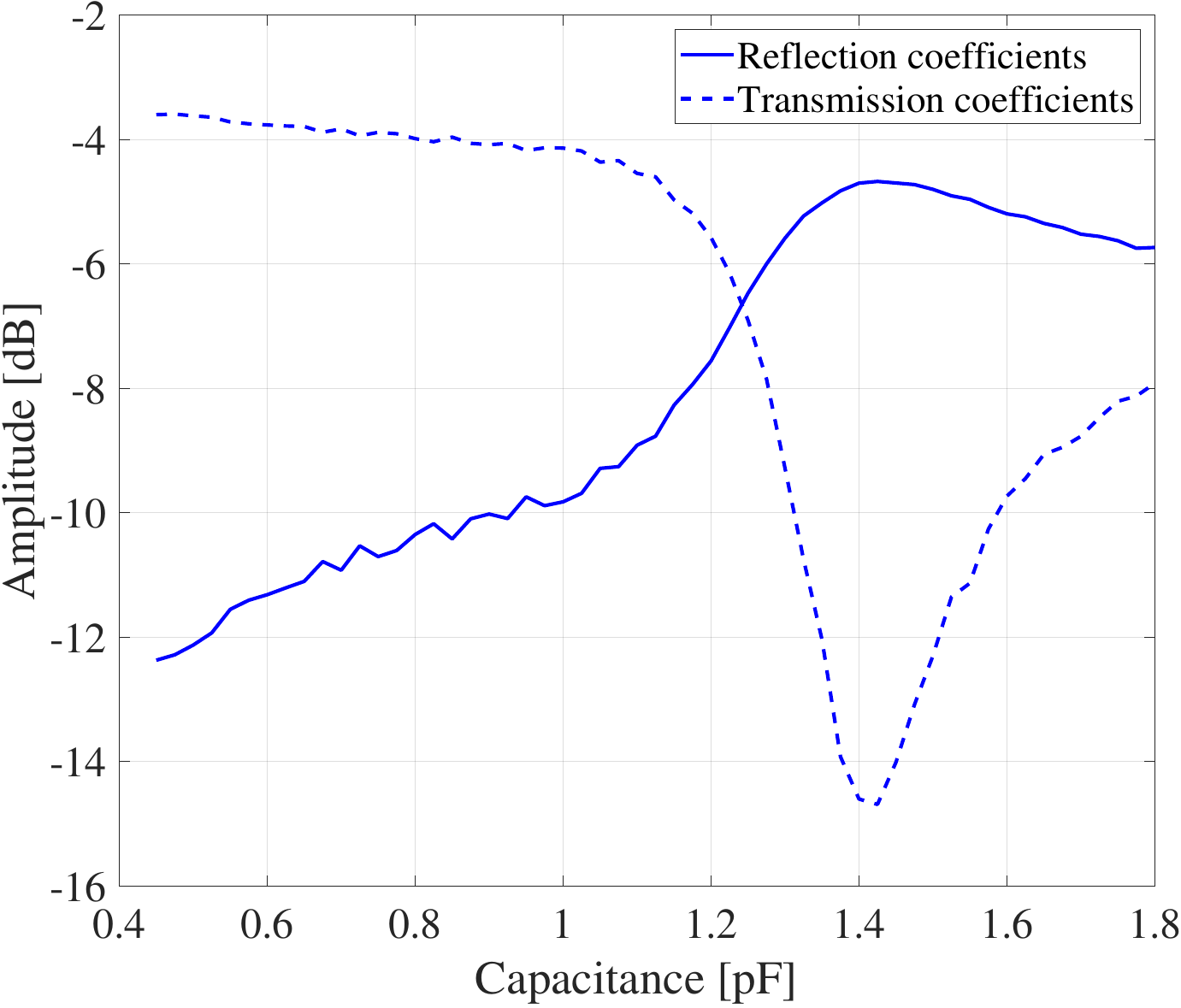}\label{fig:Diode_2G6}}
    \caption{Simulation results of the STAR-RIS unit: (a) Reflection coefficients ($S_{11}$) under different capacitance values of varactor diodes. (b) Amplitude response of reflection ($S_{21}$) and transmission ($S_{31}$) modes under different capacitance values. (c) Phase response of the reflection mode. (d) Phase response of the transmission mode. (e) Amplitude response of reflection and transmission modes with power amplifier. (f) Amplitude response at 2.6 GHz under different capacitance values.}
    \label{fig:Results of STAR-RIS}
\end{figure}

The simulation results of the unit are shown in Fig.~\ref{fig:Results of STAR-RIS}. Fig.~\ref{fig:Unit_S11} presents the simulation results for the reflection coefficients. Across a 50 MHz bandwidth around 2.6 GHz, the reflection coefficients remain below -10 dB. As the capacitance of the varactor diodes varies (adjusted via bias voltage), the reflection coefficients exhibit minimal degradation, and the center frequency shift is negligible. This indicates that the unit design demonstrates high stability and reliability in receiving incident electromagnetic waves. Additionally, the dynamic tuning of transmission and reflection power, achieved by adjusting the capacitance, does not result in significant degradation of the reflection coefficients, further validating the robustness of the design.

To accurately assess the insertion loss of the unit, simulations were first conducted without incorporating power amplifiers, with the results shown in Fig.~\ref{fig:Unit_S21_31}. When the capacitance of the varactor diodes is set to 1.224 pF, corresponding to equal amplitude transmission and reflection outputs, the insertion losses of $S_{21}$ and $S_{31}$ within the passband are both better than -7 dB. Considering the non-negligible losses introduced by the selection switch, the actual insertion loss of this structure is estimated to be less than -4 dB, demonstrating a high level of transmission efficiency.

Next, the capacitance of the varactor diodes was adjusted to 0.45 pF and 1.35 pF based on results from the evaluation board. The simulation results indicate that the STAR-RIS unit exhibits tunability of the transmission-reflection power ratio as the capacitance changes. When the capacitance is set to 0.45 pF, the insertion loss in the transmission mode is below -4.5 dB, while the reflection mode exhibits an insertion loss of approximately -9 dB. This power distribution ratio is limited by the minimum capacitance of the varactor diodes, and further optimization could be achieved by selecting diodes with even lower capacitance values.

Phase control is another critical aspect of STAR-RIS unit design, and we conducted detailed simulations to evaluate this capability. As shown in Fig.~\ref{fig:Ref_phase} and Fig.~\ref{fig:Trans_phase}, both in transmission and reflection modes, the phase control at 0° and 180° remains stable as the power distribution ratio changes, with phase errors consistently within 10°. These results demonstrate that the design maintains a high degree of phase control precision, even under varying power distribution conditions, thereby enhancing its practicality and effectiveness in real-world applications.

Finally, we evaluated the enhancement in unit performance resulting from the integration of a power amplification circuit, with the simulation results presented in Fig.~\ref{fig:PA_S21_31}. As expected, the power amplification circuit improves the gain, with an increase of more than 13 dB compared to the amplitude response in Fig.~\ref{fig:Unit_S21_31}, which did not include the amplifier. Additionally, we further adjusted the capacitance values of the varactor diodes to 0.45 pF, 1.224 pF, and 1.35 pF, observing that the variation trends and magnitudes of the transmission and reflection amplitudes closely matched those in Fig.~\ref{fig:Unit_S21_31}. The simulation results also confirmed the excellent performance of the unit design in dynamically controlling the transmission-reflection power ratio in Fig.~\ref{fig:Diode_2G6}. These findings demonstrate that the inclusion of the power amplification circuit not only significantly enhances gain but also allows for effective power distribution control under different capacitance conditions, further improving the tunability and application flexibility of the unit. During the simulation analysis of $S_{21}$ and $S_{31}$ parameters, transmission zeros were observed at frequencies of 2.5 GHz and 2.7 GHz on either side of the central frequency band. This outcome effectively confirms the capability of the system design to suppress out-of-band harmonic.

\subsection{Duplex Performance of the STAR-RIS Unit}
\label{subsec:Unit Duplex}

In the design of active RIS systems, achieving full-duplex (simultaneous bidirectional communication) functionality faces several key challenges primarily due to the design characteristics and structural limitations of current radio frequency amplification circuits. Most amplifiers are optimized for unidirectional signal amplification and are designed with high isolation to ensure effective signal boosting in one direction. However, this design approach is inadequate for full-duplex systems, which require the simultaneous handling of both forward and reverse signals. Traditional active RIS designs often struggle to balance efficient bidirectional signal processing with effective amplification, thereby limiting the implementation of full-duplex functionality. In this design, we seek to address this challenge by developing solutions that enhance the full-duplex functionality of active STAR-RIS systems.

\begin{figure}[!htbp]
    \centering 
    \subfigure[]{
    \includegraphics[width=0.3\textwidth]{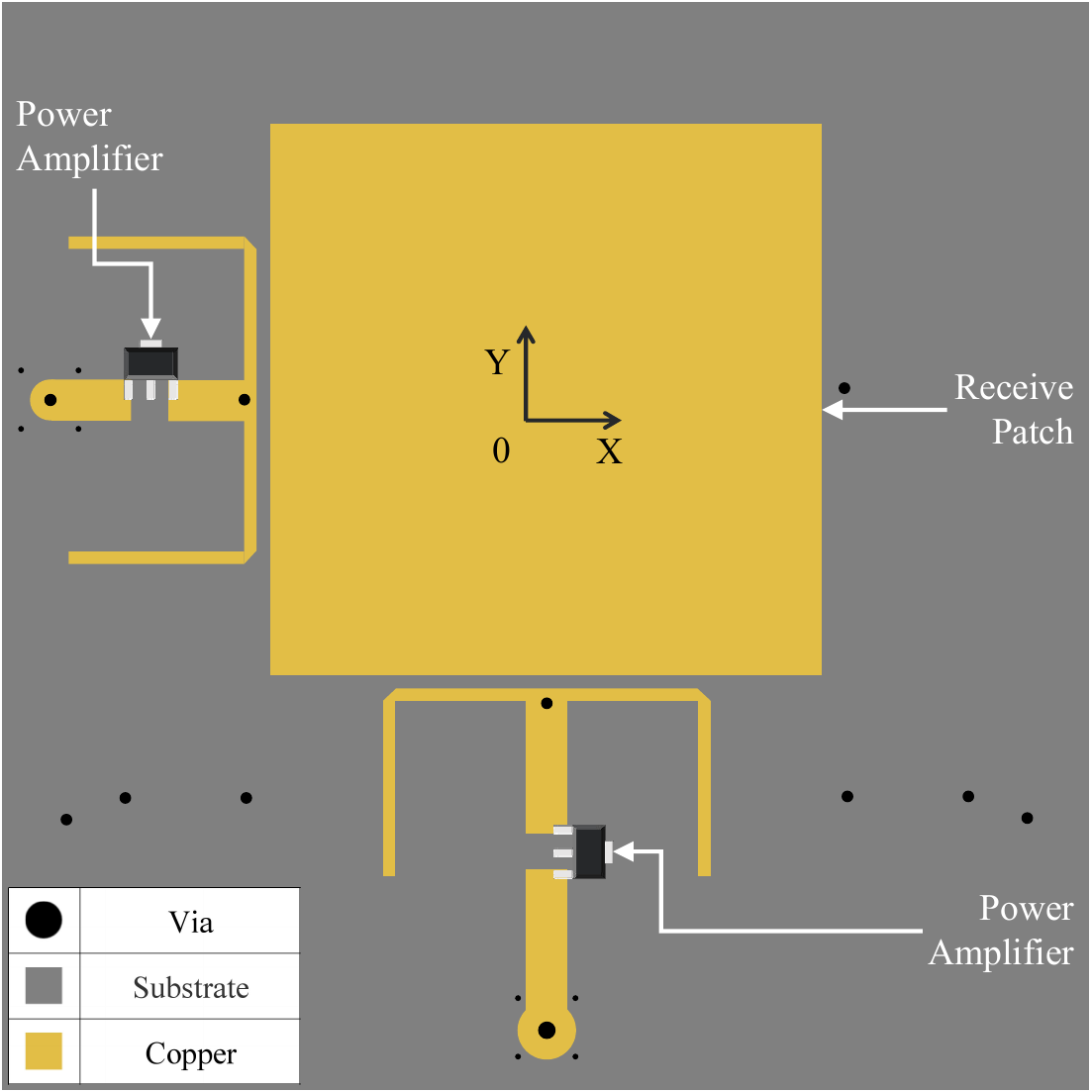}\label{fig:Duplex_unit}}
     \subfigure[]{
    \includegraphics[width=0.3\textwidth]{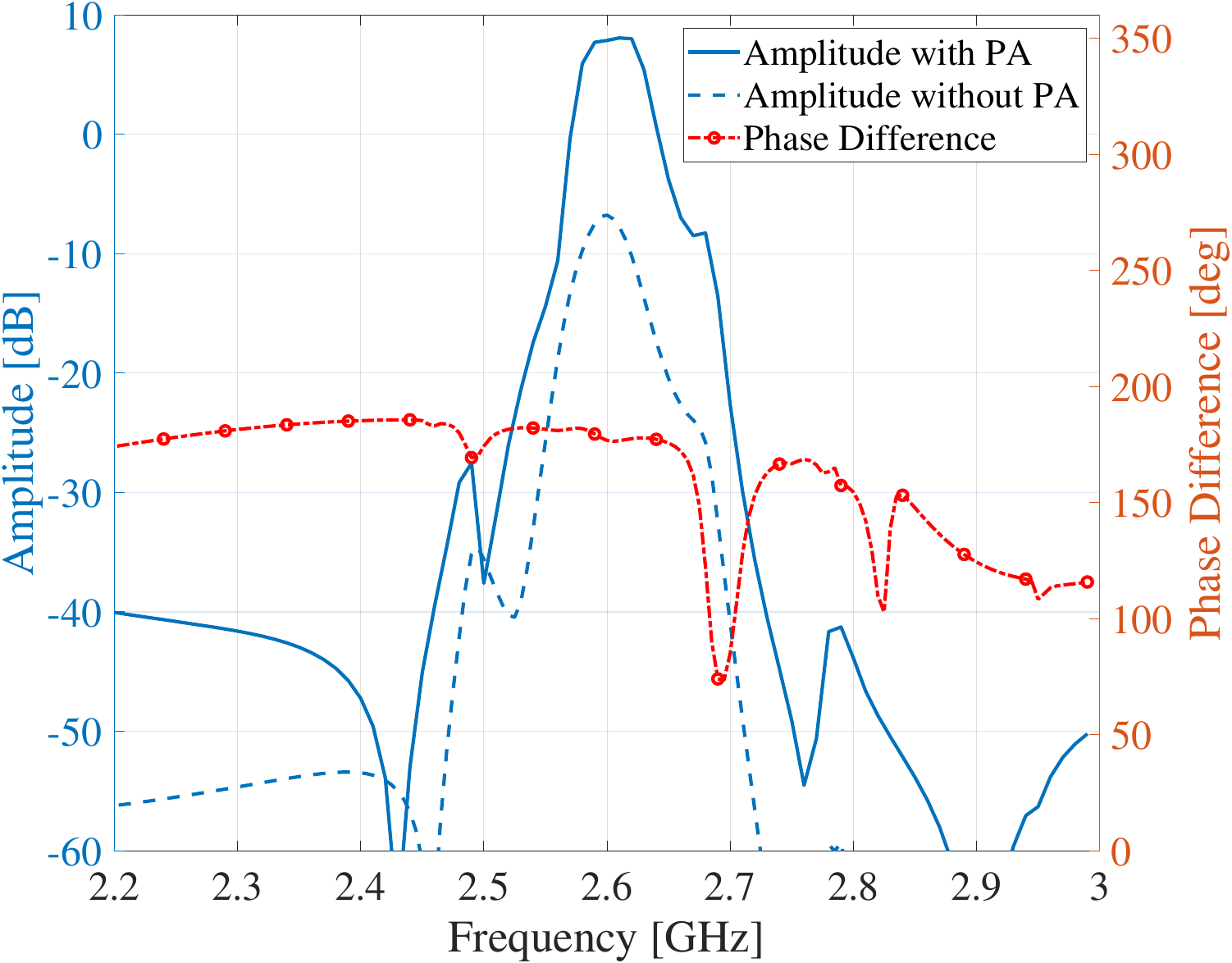}\label{fig:Duplex_result}}
    \caption{Duplex STAR-RIS unit structure and simulation results: (a) Top view of the duplex STAR-RIS unit. (b) Simulation results of the duplex STAR-RIS unit for reverse transmission.}
    \label{fig:Duplex STAR-RIS}
\end{figure}

The full-duplex STAR-RIS structure is shown in Fig.~\ref{fig:Duplex_unit}. When a unit needs to perform reverse transmission, the capacitance of the varactor diode can be adjusted by tuning the voltage, setting the capacitance to its maximum value of 7.2pF, thus switching the unit into reverse transmission mode. The simulation results of this unit are shown in Fig.~\ref{fig:Duplex_result}, demonstrating that the insertion loss for reverse transmission is better than -7dB. Compared to previous designs, this performance degradation is mainly due to the design of the adjustable power splitter. To ensure high total efficiency in both reflection and transmission for forward transmission (i.e., efficiency $\eta_n$ as described in Eq.~\ref{eq:Divider}), it is necessary to maintain sufficient isolation between port 2 and port 3 in Fig.~\ref{fig:E-board}. This isolation is a higher priority in the design of our STAR-RIS unit; therefore, despite with capacitance of the varactor diode adjusted to 7.2 pF, the insertion loss remains relatively high. However, compared to existing transmissive RIS structures, the insertion loss better than -7dB is still acceptable.

To meet the system gain requirements for reverse transmission, we add a power amplifier to the port originally used for forward signal reflection (as shown in Fig.~\ref{fig:Duplex_unit}), which significantly enhances the signal gain for reverse communication. The simulation results (Fig.~\ref{fig:Duplex_result}) validate the effectiveness of this design modification. By employing this method, we further optimize the reverse transmission capability of the unit, particularly in scenarios with high reverse communication demands.

Furthermore, the reverse transmission still supports 1-bit phase control. Unlike the previous structure, the signal, in this case, must pass through two phase control circuits sequentially. By fixing the switching state of one phase control circuit and adjusting the switches of the remaining one, different paths can be selected to achieve either $0^\circ$ or $180^\circ$ phase output. The simulation results (Fig.~\ref{fig:Duplex_result}) also confirm the feasibility of this design.

It is important to note that this is an initial design aiming to address the gap in full-duplex communication within active RIS  systems. To the best of our knowledge, this is also the first effective proposal for a full-duplex active RIS design, particularly for active STAR-RIS. This design demonstrates the feasibility of full-duplex communication in active RISs.

\section{Measurements and Verification}

In this section, we first provide a detailed description of the design of a STAR-RIS array prototype. Subsequently, we conduct a series of measurements on an \( 8 \times 4 \) STAR-RIS array, formed by combining two of the aforementioned 16-element arrays, in order to thoroughly evaluate and validate the performance of the proposed STAR-RIS design. Finally, we perform experimental validation of the STAR-RIS path loss model, derived from the RCS-based approach introduced in Sec. \ref{subsec:RCS}, using the same measurement platform.

\subsection{Design of the STAR-RIS Array Prototype}
\label{subsec:STAR-RIS Array Prototype}

\begin{figure}[!htbp]
    \centering 
    \subfigure[]{
    \includegraphics[width=0.2\textwidth]{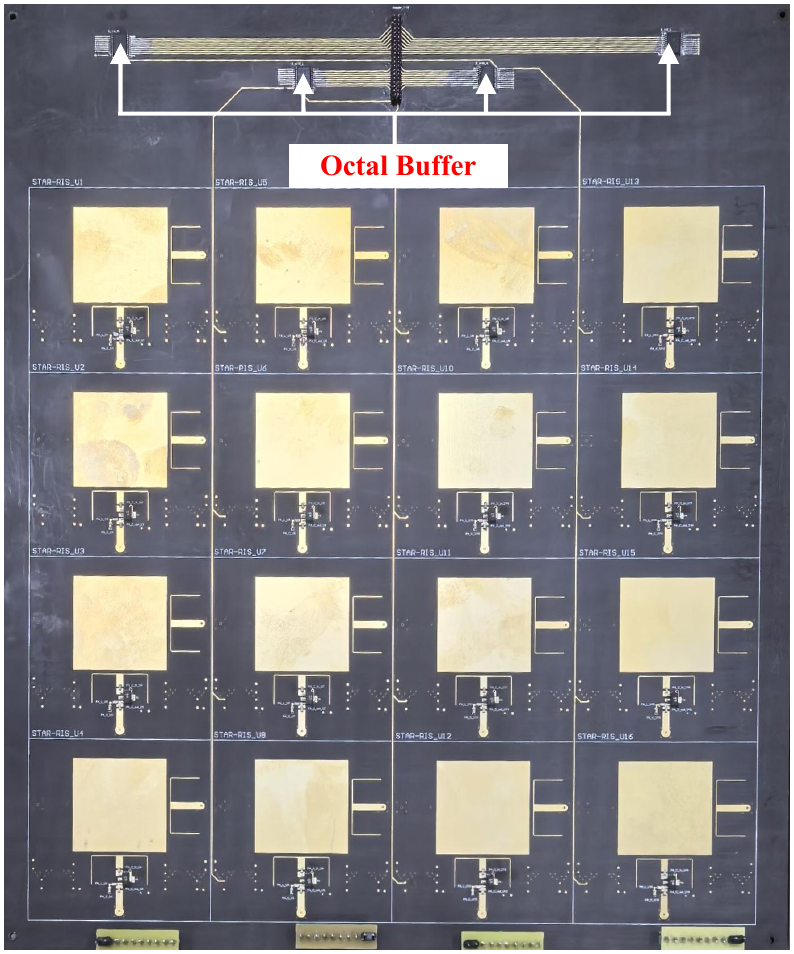}\label{fig:ARRAY_TOP}}
     \subfigure[]{
    \includegraphics[width=0.2\textwidth]{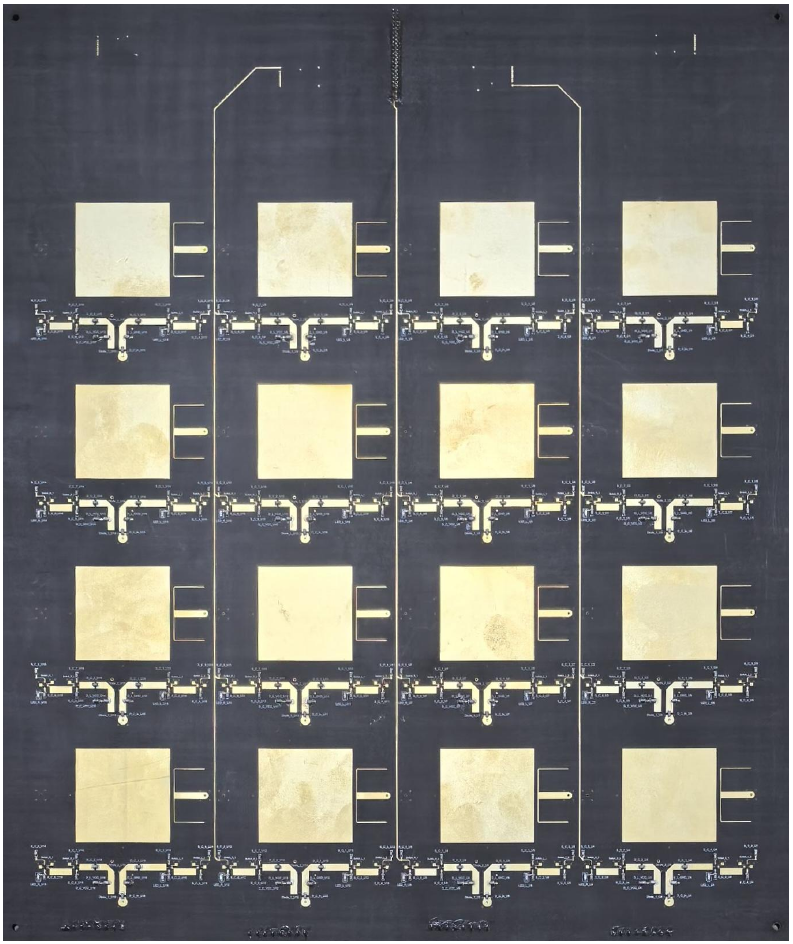}\label{fig:ARRAY_BOTTOM}}
    \caption{Schematic diagram of a 4x4 STAR-RIS array structure: (a) Top view of the STAR-RIS array. (b) Bottom view of the STAR-RIS array.}
    \label{fig:STAR-RIS_ARRAY}
\end{figure}

The design of the STAR-RIS array prototype is illustrated in Fig. \ref{fig:STAR-RIS_ARRAY}. The \( 4 \times 4 \) STAR-RIS array consists of 16 STAR-RIS units arranged in a tightly packed configuration, as detailed in Sec. \ref{subsec:Unit design}. Each phase-shifting circuit within these units is composed of two symmetrically positioned selection switches. As a result, for any given phase state, the control signals of any two selection switches are always complementary: one switch is driven by a low-level signal (logic 0, with voltage in the range of 0 to 0.3 V), and the other by a high-level signal (logic 1, with a voltage of 2.8 V). This complementary behavior ensures that the switches are configured to operate in opposite states, which is crucial for the proper functioning of the STAR-RIS array.

To reduce the complexity of the control circuitry and minimize the communication overhead between the array and the control board, 4 Octal Buffers/Drivers are deployed at the top of the STAR-RIS array. These Buffers/Drivers generate 32 independent inverted control signals, which provide the necessary phase-shifting control for each unit in the array. By using these buffers, it is possible to independently control the transmission and reflection phase states of each STAR-RIS unit, allowing for more flexible and precise manipulation of the signal propagation.

Compared to a configuration that does not employ the Octal Buffers/Drivers, the number of signal outputs required from the control board is significantly reduced. Specifically, the signal outputs are decreased from 64 to 32, resulting in a 50\% reduction in the number of control signals needed. This reduction not only simplifies the design and operation of the control system but also lowers the communication cost between the control board and the array. Despite this reduction in signal outputs, the system still ensures the independent control of both the transmission and reflection phases for each unit, maintaining the desired performance and functionality of the STAR-RIS array.

The row-pin interface located at the center of the top of the array facilitates communication with the control board, primarily handling phase selection, power supply to the Octal Buffers/Drivers, and powering the selection switches. At the bottom of the array, four additional row-pin interfaces are provided to independently supply power to the power amplifier chips and the adjustable power distribution circuits for each unit. This design enables independent control over the active gain of each unit, as well as the power distribution of the transmitted and reflected signals. In addition, 32 light-emitting diodes (LEDs) are deployed on the bottom surface of the array to display the real-time transmission and reflection phases of each unit.

\subsection{Measurement Results of the 8 $\times$ 4 STAR-RIS Array}
\label{subsec:STAR-RIS Array Measurement}

\begin{figure}[!htbp]
    \centering 
    \subfigure[]{
    \includegraphics[width=0.3\textwidth]{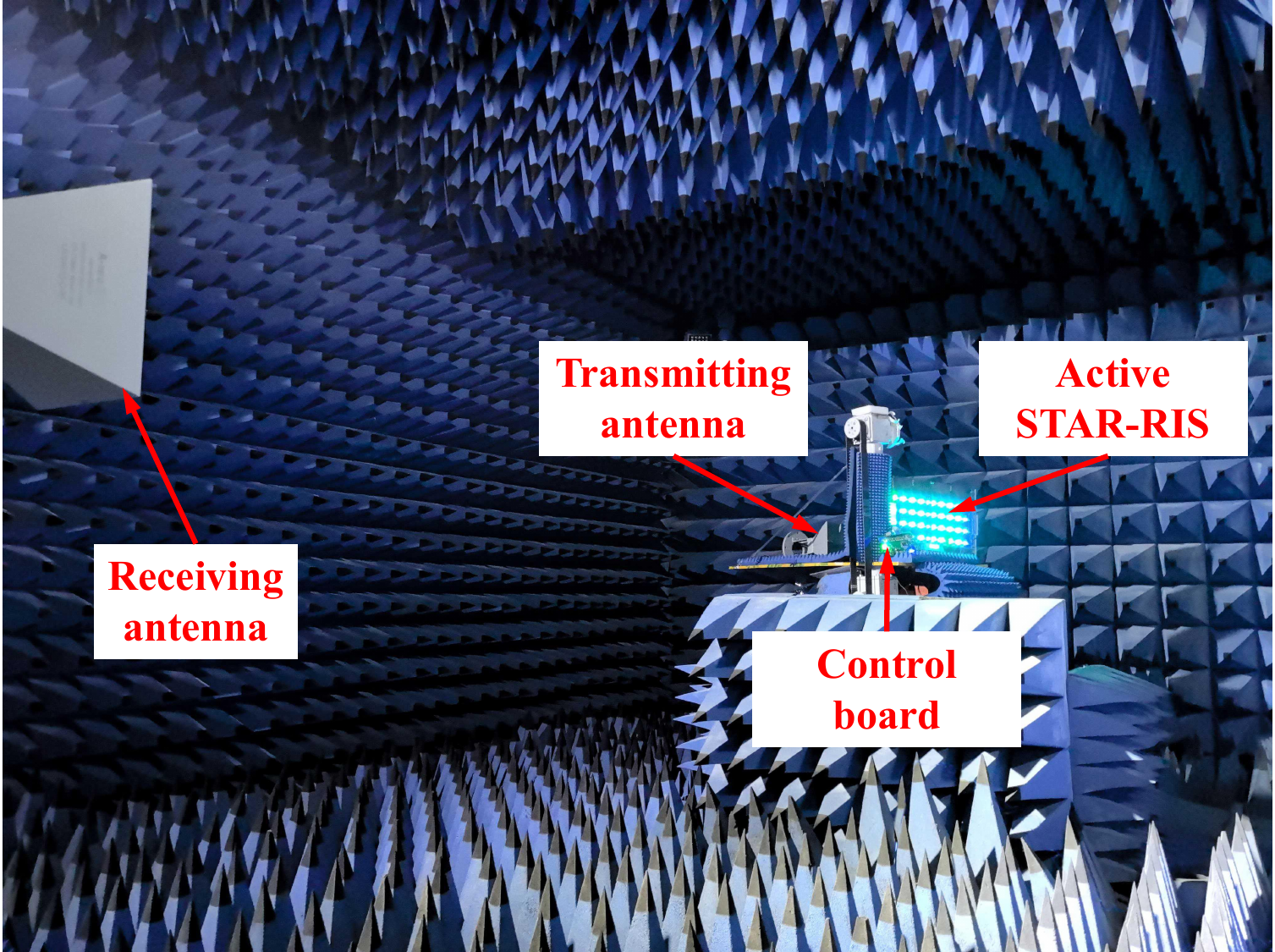}}
    \caption{STAR-RIS array measurement environment.}
    \label{fig:Environment}
\end{figure}

To conduct a more comprehensive performance evaluation of the STAR-RIS array, measurements were performed in a microwave anechoic chamber on an 8 $\times$ 4 RIS array, which was constructed by combining two 4 $\times$ 4 arrays. The measurement setup is illustrated in Fig. \ref{fig:Environment}. The STAR-RIS array is precisely positioned at the center of a turntable and controlled by a control board to facilitate real-time phase state adjustments. In the experimental setup, two horn antennas with orthogonal polarization directions are used as the transmitter and receiver, respectively, placed within the measurement area of the array. The transmitter is vertically aligned with the incident surface of the array, maintaining an 80 cm distance to generate the incident wave, ensuring uniform coverage across the entire surface of the array. The receiver is positioned 4 meters away from the STAR-RIS array at the far end of the microwave anechoic chamber to capture the electromagnetic waves transmitted or reflected by the array. This setup enables the evaluation of performance in terms of signal transmission characteristics under various operating conditions.

\subsubsection{Characterization of Phase and Amplitude Control Performance}

We conducted tests to evaluate the active gain control capabilities of the STAR-RIS array. It should be noted that, in the experimental setup described above, measuring the reflected electromagnetic waves became more challenging due to the positioning of the transmitter between the receiver and the RIS array. This issue becomes especially significant when the transmitter, receiver, and RIS array are aligned on the same straight line. As a result, accurately measuring and assessing the control of reflected signals by the RIS array presents certain challenges, especially during reflection signal amplitude tests. For this reason, during amplitude measurements, we primarily focused on the controllability of amplitude in transmission scenarios and combined the results from the evaluation board tests to verify the overall control performance of the array.

\begin{figure}[!htbp]
    \centering 
    \subfigure[]{
    \includegraphics[width=0.3\textwidth]{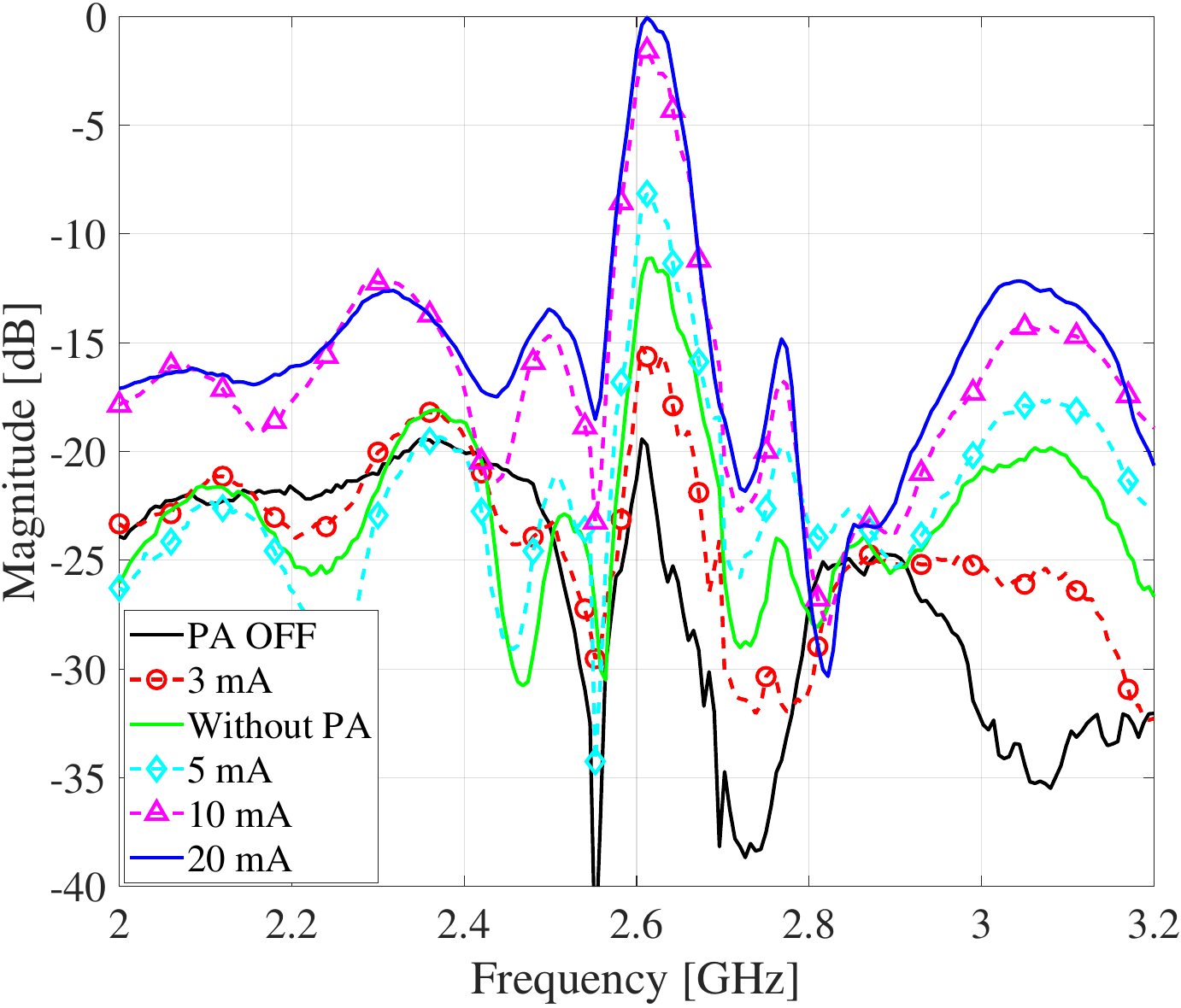}\label{fig:ARRAY_PA_STAGE}}
     \subfigure[]{
    \includegraphics[width=0.3\textwidth]{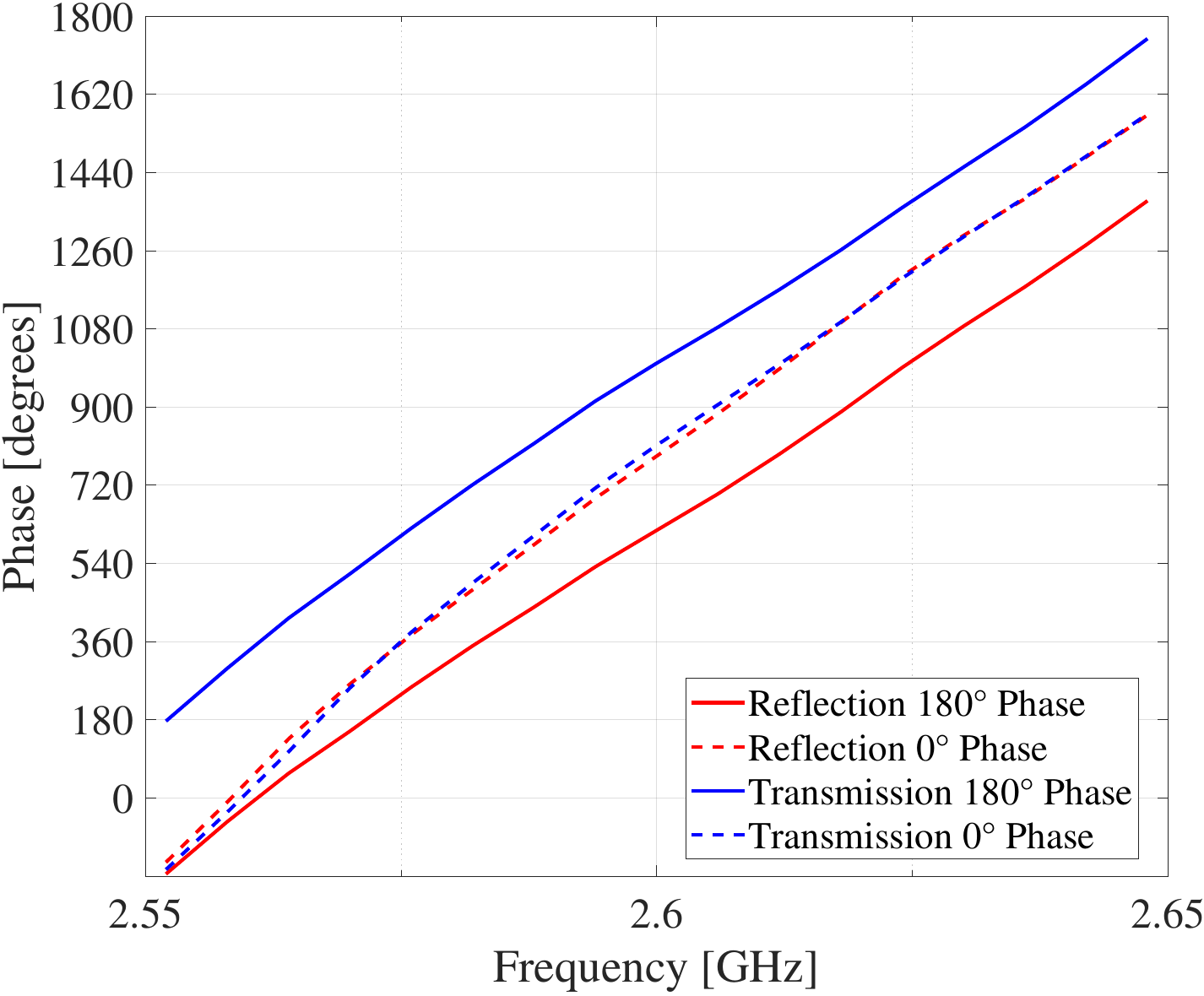}\label{fig:ARRAY_phase}}
    \subfigure[]{
    \includegraphics[width=0.3\textwidth]{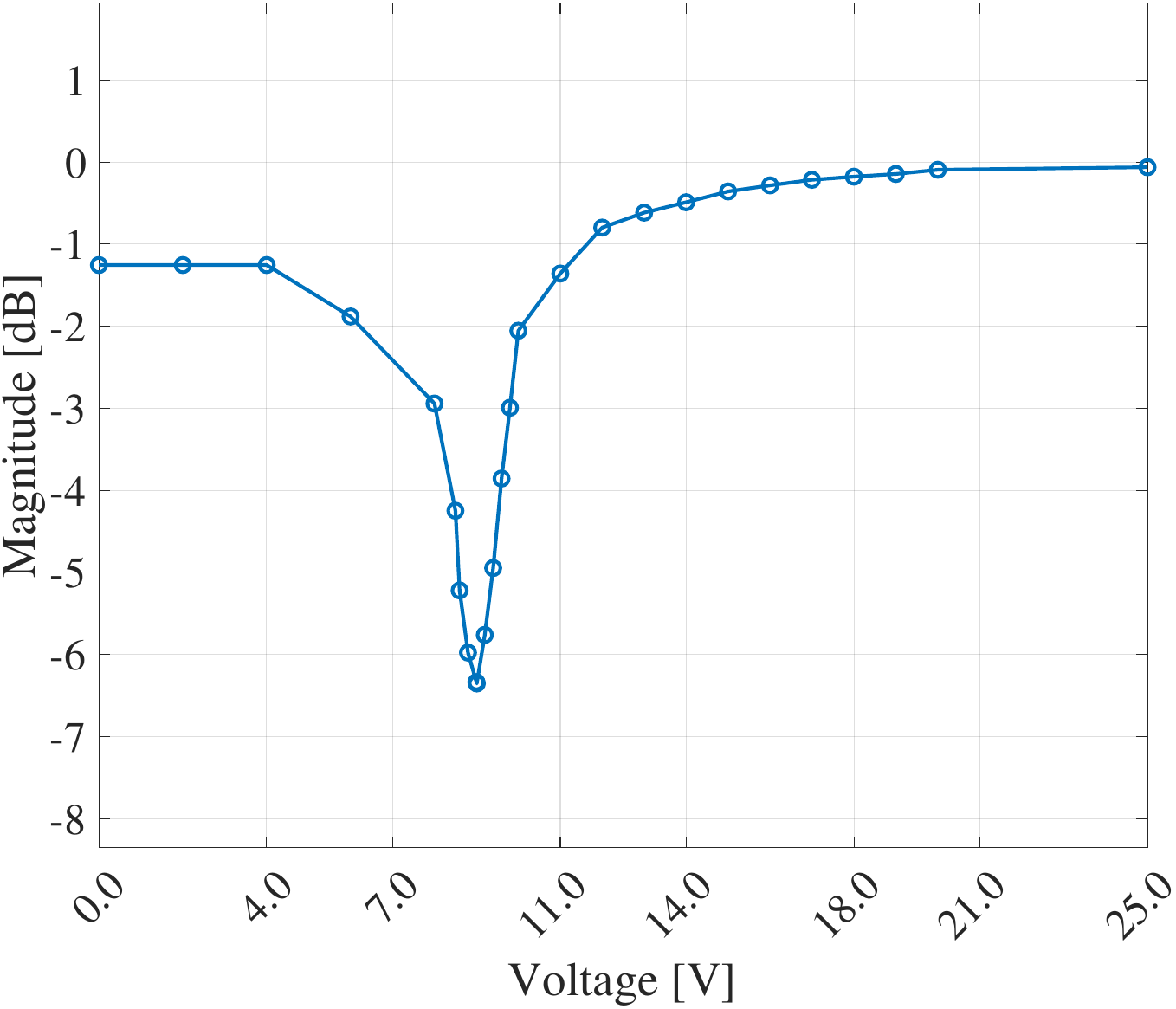}\label{fig:Array_Diode}}
    \caption{Experimental measurement results of the STAR-RIS array: (a) Measurement of the active gain of the STAR-RIS array. (b) Measurement of the phase control characteristics of the STAR-RIS array. (c) Dynamic power control of transmitted electromagnetic waves through the STAR-RIS array.}
    \label{fig:ARRAY_measurement}
\end{figure}

First, we measured the active gain of the STAR-RIS array, focusing on how the active gain provided by the power amplifier varies with the PA current (i.e., operating power) in a transmission environment. In this measurement, the transmitter and receiver were positioned on opposite sides of the RIS array, with all three devices aligned along the same straight line, resulting in both the angle of incidence and the angle of transmission being 0°($\theta_{r_t} = \theta_{r_r} = \theta_{inc} = 0^\circ$). In this configuration, the transmitted wave is incident perpendicular to the RIS array, and the receiver directly receives the transmitted wave. The measurement results are shown in Fig. \ref{fig:ARRAY_PA_STAGE}. In the measurement results, the gain of the RIS array without the power amplifier is relatively low. However, when the PA was introduced, the active gain of the RIS array increased significantly. As the PA current was increased, the gain progressively exceeded the baseline value observed without the PA, demonstrating a positive correlation with the PA current. At a PA current of 5 mA, the power amplifier provided approximately 4 dB of active gain compared to the baseline, and at 10 mA, the active gain reached approximately 10 dB. As the PA current continued to increase, the gain continued to grow, but at a diminishing rate. When the PA current reached 20 mA, which corresponds to the maximum operating current of the power amplifier, the RIS array gain saturated, stabilizing around 12 dB. This saturation indicates that the gain reached its maximum value, and further increases in current did not result in a significant gain improvement. These measurement results align with those obtained from the evaluation board, validating the performance of the active STAR-RIS design in providing adjustable active gain.

Next, we conducted measurements to assess the phase-shifting functionality of the STAR-RIS array. During the phase measurement process, the transmission phase control capability is first assessed according to the previously described setup. Then, the central rotating platform is rotated by 180°, maintaining the alignment of the receiver, transmitter, and RIS array along the same horizontal plane. However, in this configuration, both the receiver and transmitter are positioned on the same side of the RIS array to measure the reflection phase control capability. In both of these configurations, we sequentially switch the control signals of the transmission and reflection selection switches to record the phase data under different control signal settings. The results are presented in Fig. \ref{fig:ARRAY_phase}. The measurement results indicate that, upon switching the control signals via the control panel, the phase difference between the transmitted and reflected electromagnetic waves reached 180°. Furthermore, within the operating bandwidth, the phase error remained consistently below 10°. These findings demonstrate the precise phase control capability of the system, highlighting its effectiveness in maintaining accurate phase alignment across the frequency range.

\subsubsection{Measurement of Dynamic Control of Electromagnetic Wave Power}

Finally, we conducted measurements to assess the dynamic control of the transmitted electromagnetic wave power, with the experimental setup being consistent with that used for the active gain measurement of the power amplifier. When the supply current to the power amplifier chip was set to 20 mA (corresponding to the maximum operating power of the power amplifier), we gradually adjusted the reverse bias voltage applied across the varactor diode and recorded the received power at the center frequency to evaluate the variation in the transmitted electromagnetic wave power. The transmitted amplitude demonstrated a dynamic control range exceeding 6 dB, with its variation trend closely matching the simulation results and evaluation measurements. This indicates that the achieved transmission amplitude adjustment aligns with theoretical expectations and prior simulation analyses, further validating the feasibility and effectiveness of the proposed scheme. Overall, the transmission amplitude exhibited smooth and stable changes during the adjustment process, in line with the anticipated design objectives.

\subsection{STAR-RIS Beamforming Performance Assessment}

In this section, we conducted a preliminary assessment of the beamforming capabilities of the 8 $\times$ 4 STAR-RIS array to verify its performance in steering beams at various angles. The testing environment remained consistent with the setup described in Sec. \ref{subsec:STAR-RIS Array Measurement}, ensuring consistency and reproducibility of the tests. To achieve efficient beamforming, we implemented a greedy algorithm previously developed by our team for RIS systems \cite{Pei2021Ref}. This algorithm iteratively adjusts the phase states of each unit in the array to form the desired beam shape at the specified steering angle.

A notable difference in this testing setup was the addition of a spectrum analyzer, which was used to monitor the received power at the receiver continuously. This setup allowed for real-time communication between the spectrum analyzer and the STAR-RIS array control board, enabling precise control of the phase states of individual array elements. Therefore, the STAR-RIS array was able to perform beamforming at any desired steering angle. The implementation of this configuration facilitated fine-tuned control of the beam direction, further validating the ability of the array to adapt to different beamforming scenarios.

Once the algorithm completed the beamforming process, the output signal from the receiver was switched to a vector network analyzer (VNA) for radiation pattern measurement. This transition allowed for the assessment of the array beamforming performance by evaluating the radiation pattern across the array operating frequency range. The results provided a comprehensive evaluation of the algorithm effectiveness and the practical performance of the STAR-RIS array. Through these experiments, we not only validated the feasibility of the proposed algorithm but also explored the impact of phase control on the array beam directionality and gain.

\begin{figure}[!htbp]
    \centering 
    \subfigure[]{
    \includegraphics[width=0.3\textwidth]{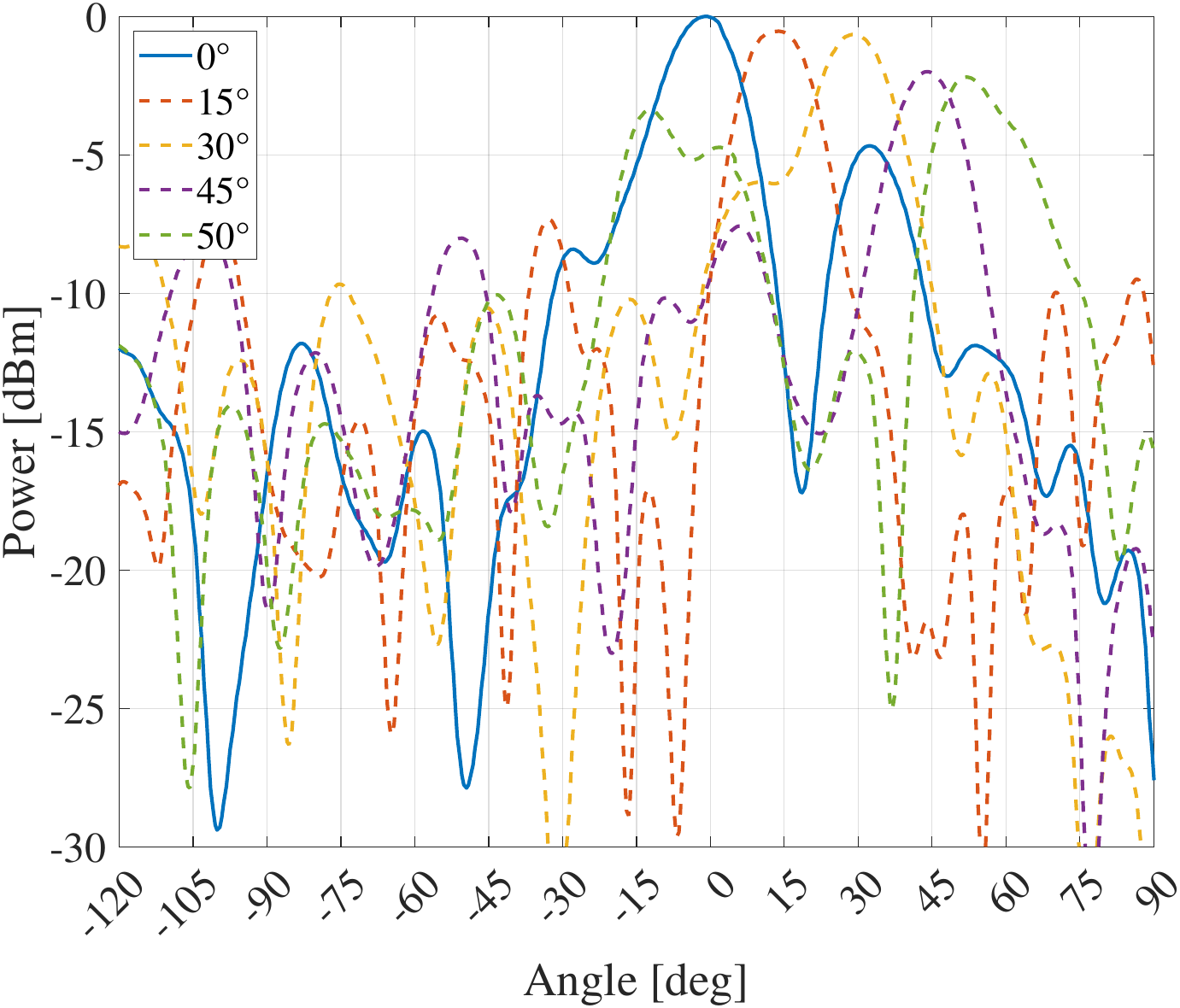}\label{fig:Trans}}
     \subfigure[]{
    \includegraphics[width=0.3\textwidth]{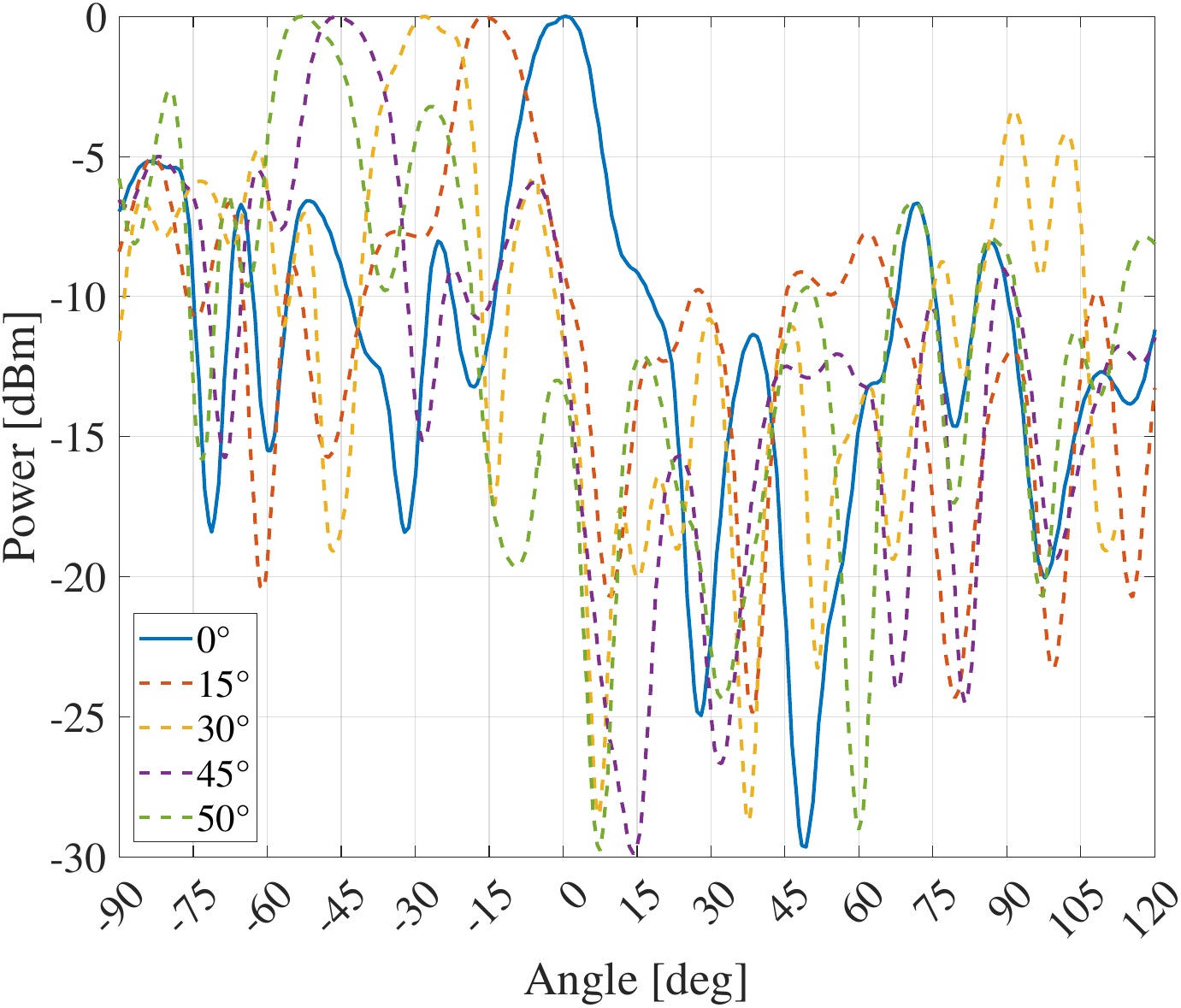}\label{fig:Ref}}
    \caption{Beamforming Characteristics of an 8 $\times$ 4 STAR-RIS Array: (a) Transmission Beamforming Pattern. (b) Reflection Beamforming Pattern.}
    \label{fig:Rediation}
\end{figure}

\subsubsection{Beamforming Performance of Transmitted Beam}

We performed measurements to assess the beamforming performance of the transmitted beam at various steering angles. The measurements started from 0° and were incremented by 15°, with the experimental results shown in Fig. \ref{fig:Trans}. For \(\theta_{r_t} = 0^\circ\), the 3dB beamwidth of the main lobe was 16°, demonstrating good beam directionality. As the steering angle increased, the beam amplitude decreased by approximately 2 dB, and at \(\theta_{r_t} = 45^\circ\), the amplitude decreased by about 2 dB, further highlighting the degradation in beam directionality and amplitude control at larger steering angles.

At \(\theta_{r_t} = 15^\circ\), the 3dB beamwidth of the main lobe did not show significant changes, with an amplitude decrease of around 0.5 dB, indicating minimal loss in beam directionality. As the steering angle was further increased to 30°, the main lobe beamwidth expanded to 17°, and the amplitude dropped by approximately 1 dB. Both beam directionality and amplitude exhibited a linear degradation, indicating that while beam directionality and gain gradually weakened with increasing steering angle, beamforming remained effective within this range.

When the steering angle increased to 45°, the amplitude decreased by approximately 2 dB, further emphasizing the loss of beam directionality and amplitude control at larger steering angles. As the steering angle increased further to 52°, the amplitude dropped by approximately 2.5 dB, and the main lobe beamwidth showed significant degradation. This suggests that beamforming performance is significantly affected at large steering angles.

Overall, although beam directionality and amplitude degrade as the steering angle increases, the STAR-RIS array still effectively controls beam directionality and maintains relatively stable amplitude at smaller steering angles. This confirms the beamforming capability of the array at moderate to small steering angles while indicating that beamforming stability decreases as the steering angle grows.

\subsubsection{Beamforming Performance of Reflected Beam}

Next, we measured the beamforming performance of the reflected beam. As mentioned earlier, the amplitude of the reflected signal is influenced by the mutual interaction between the TX and RX when they are positioned on the same side. To minimize this interference, we changed the transmitter position to a 45° oblique incidence (\(\theta_{inc} = 45^\circ\)) for the test. Since this setup differs from the one used for the transmission tests, we normalized the amplitude of all reflected beams in this section, setting the maximum power point to 0 dB. The test results are shown in Fig. \ref{fig:Ref}.

The results indicate that the beamforming performance of the reflected beam is similar to the transmission mode. In the range of 0° to 45° (\(\theta_{r_r}\)), the 3dB beamwidth of the main lobe did not show significant degradation. At \(\theta_{r_r} = 0^\circ\), the 3dB beamwidth was approximately 16°, while at \(\theta_{r_r} = 45^\circ\), it increased to around 20°, demonstrating good beam control. However, when the steering angle increased to 50°, the performance showed a noticeable decline, with the beamwidth increasing and the amplitude decreasing, indicating that as the steering angle increases, the beamforming performance of the reflected beam deteriorates.

These results further validate the beamforming capability of the reflected beam, showing that within smaller steering angle ranges, the reflected beam performs similarly to the transmitted beam, maintaining good beam directionality. However, as the steering angle increases, the stability of the beamforming performance begins to decrease, providing valuable insights for further optimization of reflected beamforming.

 \subsection{Validation of the STAR-RIS Path Loss Model based on RCS}
 \label{subsec:STAR-RIS Array RCS}

Furthermore, we validated the RCS-based model proposed and extended it to STAR-RISs in Sec. \ref{subsec:RCS}. In these tests, the transmitter was positioned 2 m from the center of the STAR-RIS array in an incident configuration (\( t_t = 2 \, \text{m}, \theta_{\text{inc}} = 0^\circ \)). In comparison, the receiver was also placed at a distance of 2 m from the array (\( r_t = r_r = 2 \, \text{m} \)). The received power was recorded at beamformed zenith deflection angles, measured at intervals of \( 10^\circ \) across a range from \( 0^\circ \) to \( 60^\circ \). Additionally, the azimuth deflection angle was maintained at a constant value of 0° ($\varphi_{r_t}=\varphi_{r_r}=0^\circ$) during these measurements.

\begin{figure}[!htbp]
    \centering 
    \subfigure[]{
    \includegraphics[width=0.3\textwidth]{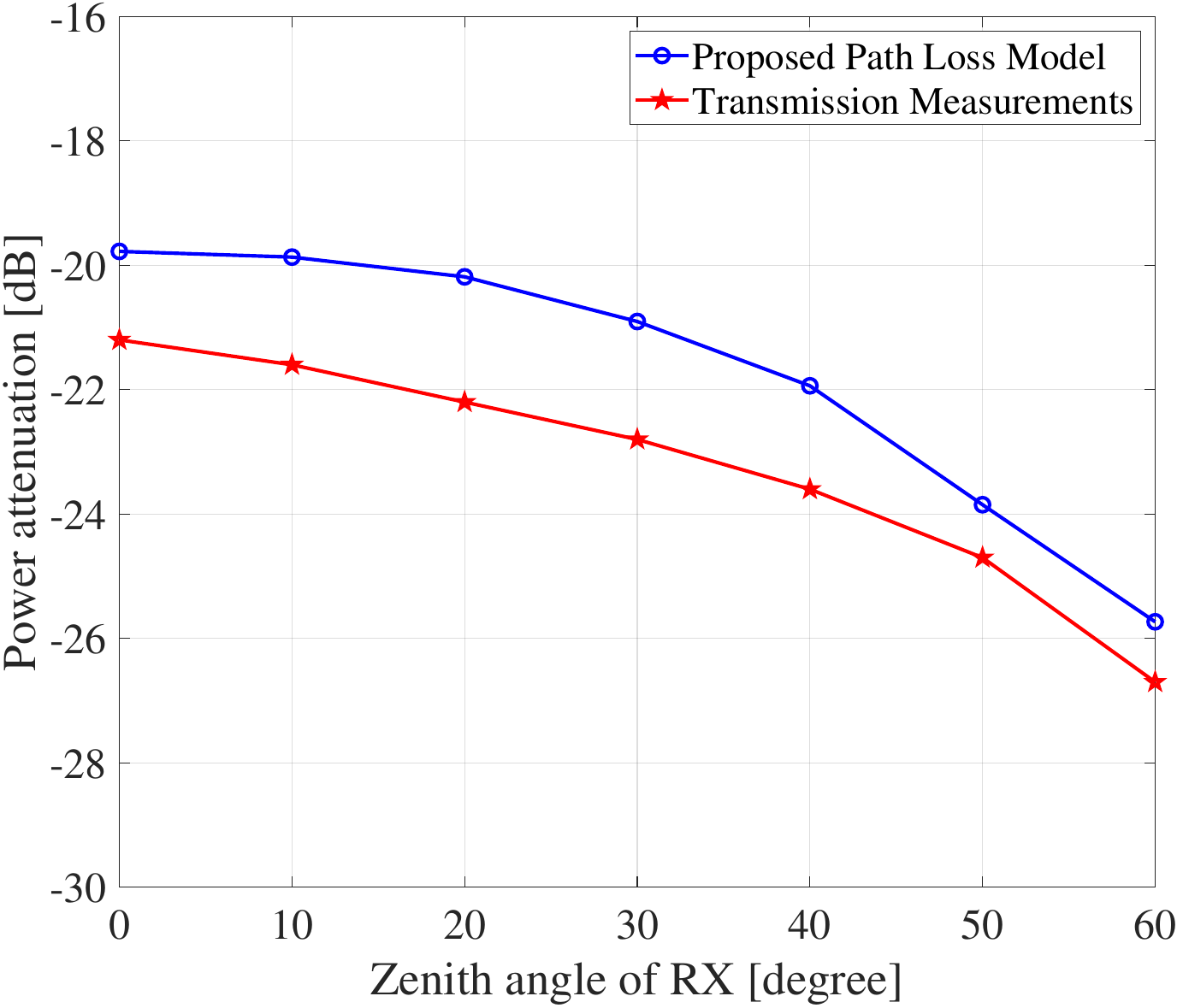}\label{fig:RCS_Trans}}
     \subfigure[]{
    \includegraphics[width=0.3\textwidth]{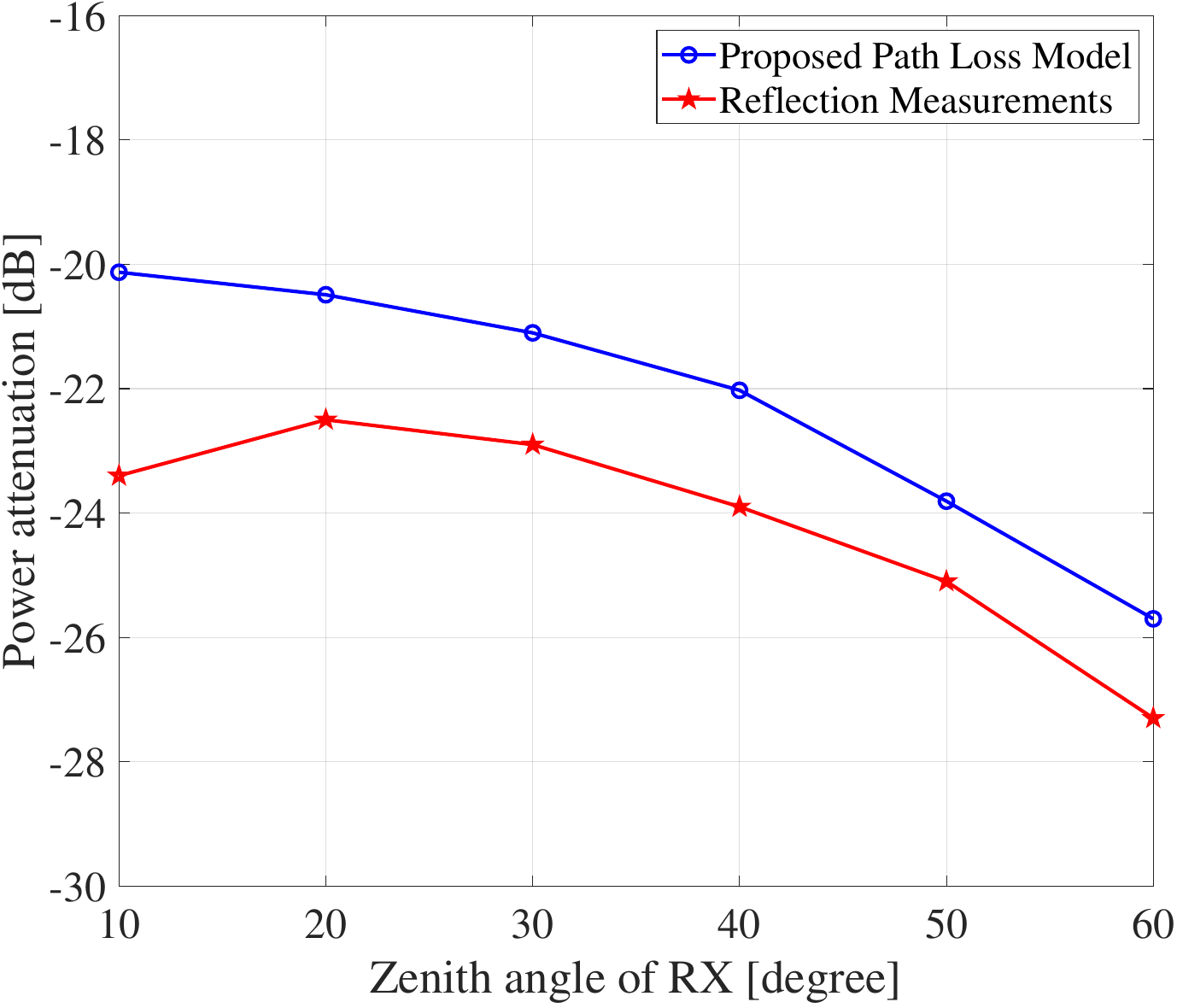}\label{fig:RCS_Ref}}
    \caption{Simulation and Measurement Results of the RCS-Based STAR-RIS Path Loss Model:
(a) Verification under different steering angles in transmission mode.
(b) Verification under different steering angles in reflection mode.}
    \label{fig:RCS_Measurement}
\end{figure}

The simulation and measurement results are shown in Fig. \ref{fig:RCS_Measurement}. The results under transmission conditions are presented in Fig. \ref{fig:RCS_Trans}. For all seven preset deflection angles, the simulation and measurement results exhibit good consistency, with the maximum error being less than 2 dB, confirming the accuracy and reliability of the proposed model under transmission conditions. The simulation and measurement results under reflection conditions are shown in Fig. \ref{fig:RCS_Ref}. It is important to note that during the validation of the reflection path loss model when \( \theta_{r_r} = 0^\circ \), the positions of the transmitter and receiver overlap. To avoid potential interference with the measurement results, the initial angle in this section of the experiments was set to \( 10^\circ \). Among the six deflection angles, except for 10°, the simulation and measurement results remain well-matched, with minimal discrepancies. Specifically, for all deflection angles other than 10°, the results align with those under transmission conditions, further validating the applicability and effectiveness of the proposed path loss model in the STAR-RIS scenario. The discrepancy observed at 10° can be attributed to the relatively large aperture size of the horn antennas used as both the transmitter and receiver in the measurement setup. At a deflection angle of 10°, the distance between the transmitter and receiver is relatively short, which may lead to interference between the two antennas, thereby affecting the accuracy of the measurement results.

\section{Conclusion}

In this paper, we proposed an active STAR-RIS design that enables independent control of both transmitted and reflected beams, as well as independent adjustment of amplitude, power split ratio, and phase for each unit. Compared to existing STAR-RIS structures, this design introduces active gain. It enables dynamic control of both transmitted and reflected power split ratios, providing greater flexibility for optimizing power distribution and coverage in wireless communication networks. Additionally, we extended the RCS-based path loss model for active STAR-RIS, building on our previous work. The design also provides up to 15 dB of active gain and filtering functionality, with experimental results showing good agreement with simulations. However, the current study only provides preliminary functional validation of the proposed system, and further research is needed, particularly in beamforming algorithms, multi-beam control, and dynamic power distribution. Overall, this design addresses the losses inherent in traditional RIS systems while enhancing STAR-RIS performance by enabling amplitude control, providing new possibilities for wireless communication applications in complex environments.

\ifCLASSOPTIONcaptionsoff
  \newpage
\fi

\bibliographystyle{IEEEtran}

\end{document}